\documentclass[5p,twocolumn]{elsarticle} 
\usepackage{tikz}
\usetikzlibrary{shapes, arrows.meta, positioning}
\usepackage[utf8]{inputenc}
\usepackage[T1]{fontenc}
\usepackage{lmodern}
\usepackage{amsmath,amssymb,amsfonts}
\usepackage{graphicx}
\usepackage{multirow}
\usepackage{booktabs}
\usepackage{hyperref}
\usepackage{xcolor}

\biboptions{sort&compress}
\usepackage{adjustbox}
\usepackage{multirow}
\usepackage{mathrsfs}
\usepackage{algorithmic}
\usepackage{array}
\usepackage[caption=false,font=normalsize,labelfont=sf,textfont=sf]{subfig}
\usepackage{textcomp}

\usepackage{url}
\usepackage{verbatim}
\usepackage{graphicx}
\usepackage[utf8]{inputenc}
\usepackage[numbers]{natbib}
\usepackage[T1]{fontenc}
\usepackage{lmodern}
\usepackage{multirow}
\usepackage{booktabs}
\usepackage{hyperref}
\usepackage{xcolor}
\usepackage{adjustbox}
\usepackage{mathrsfs}
\usepackage{dblfloatfix}
\usepackage{amsmath,amssymb,amsfonts}
\usepackage{mathtools} 
\usepackage{float}
\journal{Information Fusion}

\begin{document}

\begin{frontmatter}

\title{MOMENTA: Mixture-of-Experts Over Multimodal Embeddings with Neural Temporal Aggregation for Misinformation Detection}

\author[1]{Yeganeh Abdollahinejad \fnref{fn1}}
\author[2]{Ahmad Mousavi\corref{cor1}\fnref{fn1}}
\ead{mousavi@american.edu}
\author[3]{Naeemul Hassan}
\author[4]{Kai Shu}
\author[5]{Nathalie Japkowicz}
\author[6]{Shahriar Khosravi}
\author[7]{Amir Karami}

\cortext[cor1]{Corresponding author}
\fntext[fn1]{These authors contributed equally to this work.}
\address[1]{Department of Computer Science and Mathematics, Pennsylvania State University, Harrisburg, PA, USA}
\address[2]{Department of Mathematics and Statistics, American University, Washington, DC, USA}
\address[3]{Philip Merrill College of Journalism and College of Information, University of Maryland, College Park, MD, USA}
\address[4]{Department of Computer Science, Emory University, Atlanta, GA, USA}
\address[5]{Department of Computer Science, American University, Washington, DC, USA}
\address[6]{Department of Philosophy, Georgetown University, Washington, DC, USA}
\address[7]{School of Data Science and Analytics, Kennesaw State University, Kennesaw, GA, USA}

\begin{abstract}
The widespread dissemination of multimodal content on social media has made misinformation detection increasingly challenging, as misleading narratives often arise not only from textual or visual content alone, but also from semantic inconsistencies between modalities and their evolution over time. Existing multimodal misinformation detection methods typically model cross-modal interactions statically and often show limited robustness across heterogeneous datasets, domains, and narrative settings. To address these challenges, we propose MOMENTA, a unified framework for multimodal misinformation detection that captures modality heterogeneity, cross-modal inconsistency, temporal dynamics, and cross-domain generalization within a single architecture. MOMENTA employs modality-specific mixture-of-experts modules to model diverse misinformation patterns, bidirectional co-attention to align textual and visual representations in a shared semantic space, and a discrepancy-aware branch to explicitly capture semantic disagreement between modalities. To model narrative evolution, we introduce an attention-based temporal aggregation mechanism with drift and momentum encoding over overlapping time windows, enabling the framework to capture both short-term fluctuations and longer-term trends in misinformation propagation. In addition, domain-adversarial learning and a prototype memory bank improve domain invariance and stabilize representation learning across datasets. The model is trained using a multi-objective optimization strategy that jointly enforces classification performance, cross-modal alignment, contrastive learning, temporal consistency, and domain robustness. Experiments on four benchmark datasets, \textit{Fakeddit}, \textit{MMCoVaR}, \textit{Weibo}, and \textit{XFacta}, show that MOMENTA achieves strong and consistent performance across multiple evaluation metrics, including accuracy, F1-score, AUC, and MCC. These findings suggest that integrating expert specialization, discrepancy-aware fusion, temporal reasoning, and domain-invariant representation learning provides an effective direction for real-world multimodal misinformation detection.
\end{abstract}


\end{frontmatter}
\label{sec:intro}

\section{Introduction}

The widespread adoption of social media has significantly increased the generation, transmission, and consumption of information. This has created a trade-off between knowledge aggregation and the spread of misinformation~\cite{acemoglu2010misinformation}. These platforms facilitate the transmission of veridical information, but also accelerate the spread of false or misleading content~\cite{floridi2011information}.  The extensive spread of misinformation via these large-scale technologies leads to various moral, social, and epistemic harms, for example, by undermining trust in public institutions. In the health domain, exposure to misleading social media content has been associated with reduced trust in healthcare institutions, lower confidence in professional medical guidance, and weaker adherence to recommended health behaviors~\cite{stimpson2025perceived,jmir2024healthmisinformation,infodemic_review_2022}. In social and political contexts, misinformation has also contributed to distrust in media, government information, and political institutions, further amplifying its spread~\cite{lewandowsky2023misinformation,lackey2021echos}. As these examples suggest, addressing misinformation is far from a merely technical issue; it is a social, political, and public health priority.

Misinformation detection becomes even more difficult when posts combine textual and visual content. In such cases, effective systems must not only understand each modality independently, but also reason over their semantic interactions~\cite{ECaTCH,ooc_multimodal_misinformation,rajabi2024event}. Content that appears veridical in isolation may become misleading when paired with unrelated, repurposed, or manipulated visual content. Prior work has shown that misleadingness can emerge from the mismatch between a headline and its accompanying video, underscoring that deceptive intent often arises from cross-modal inconsistency rather than from either modality alone~\cite{Naeemul}. Earlier approaches frequently processed posts independently or relied on shallow fusion strategies such as simple feature concatenation, limiting their ability to capture semantic disagreement and richer multimodal dependencies.

Recent work has increasingly explored foundation-model-based architectures for multimodal misinformation detection. Large pretrained encoders such as XLM-RoBERTa and CLIP provide strong cross-lingual and cross-modal representations for heterogeneous social media data~\cite{conneau2020xlm,radford2021learning,zhang2025kamp}. At the same time, Mixture-of-Experts (MoE) architectures have emerged as an effective way to model the heterogeneity of misinformation by routing inputs through specialized expert networks. Frameworks such as MIMoE-FND dynamically model unimodal agreement and cross-modal semantic alignment, while MisD-MoE uses modality-specific experts and adaptive feature selection to reduce redundancy and improve classification robustness~\cite{liu2025mimoe,liu2024misdmoe}. Despite these advances, most existing approaches still treat cross-modal relationships as largely static and do not explicitly model how misinformation narratives evolve over time.

Another major challenge is domain shift. Models trained on one platform or dataset often generalize poorly to new domains with different linguistic styles, visual distributions, topic structures, or cultural contexts. Recent event-centric methods such as E-CaTCH have shown that temporal information, including semantic drift and narrative progression, can substantially improve misinformation detection robustness~\cite{ECaTCH}. However, jointly integrating temporal reasoning, cross-modal alignment, and domain-invariant representation learning remains an underexplored problem.

In this work, we propose \textbf{MOMENTA}, a \textit{Mixture-of-Experts over Multimodal Embeddings with Neural Temporal Aggregation} framework for multimodal misinformation detection. MOMENTA employs modality-specific MoE layers with input-dependent gating to capture diverse misinformation patterns without introducing excessive architectural complexity. Bidirectional co-attention aligns textual and visual representations in a shared semantic space, while an auxiliary discrepancy branch explicitly models semantic inconsistencies between modalities. To represent narrative evolution, we aggregate posts through overlapping temporal windows and encode both drift and momentum. We further incorporate domain-adversarial learning~\cite{ganin2016domain} and a prototype memory bank to encourage domain-invariant representations across datasets. The model is trained using a multi-objective optimization strategy that jointly supports classification, cross-modal alignment, contrastive learning, temporal consistency, and domain robustness. Through experiments on Fakeddit, MMCoVaR, Weibo, and XFacta, we show that MOMENTA achieves strong and consistent performance across diverse multimodal misinformation benchmarks.

Our main contributions are summarized as follows:
\begin{itemize}
    \item We propose MOMENTA, a unified multimodal misinformation detection framework that combines modality-specific expert specialization, cross-modal alignment, discrepancy-aware fusion, temporal aggregation, and domain generalization within a single architecture.
    \item We introduce a temporal aggregation mechanism based on overlapping windows with drift and momentum encoding to model the evolution of misinformation narratives over time.
    \item We incorporate domain-adversarial learning and a prototype memory bank to encourage domain-invariant representations and improve robustness across heterogeneous datasets.
    \item We evaluate MOMENTA on four benchmark datasets and show strong and consistent performance across multiple multimodal misinformation detection settings.
\end{itemize}

\section{Related Work}

\subsection{Multimodal Misinformation Detection}

The increasing sophistication of misinformation in online ecosystems has rendered unimodal detection approaches insufficient. Early methods focused on either textual linguistic features or visual artifacts independently, failing to capture the complex interactions between modalities that adversaries exploit. As a result, recent research has shifted toward multimodal frameworks that jointly analyze textual and visual signals to improve detection robustness and generalization~\cite{Shang2022DGExplain,dEFEND,LEMMA}.

Modern multimodal systems typically rely on large-scale pretrained models to encode heterogeneous inputs into semantically meaningful representations. Transformer-based language models such as XLM-RoBERTa~\cite{conneau2020xlm} and vision-language models such as CLIP~\cite{radford2021learning} have become foundational due to their strong cross-lingual and cross-modal generalization capabilities. These embeddings are subsequently projected into a shared latent space to facilitate interaction across modalities. This projection step is critical, as it enables semantic alignment between heterogeneous inputs rather than serving as a simple dimensionality reduction process~\cite{LEMMA,MMfakeBench}.

Despite these advances, multimodal detection remains challenging due to the inherent semantic gap between modalities and the adversarial nature of misinformation, motivating the development of more sophisticated fusion and reasoning mechanisms~\cite{ECaTCH,XFacta}.

\subsection{Cross-Modal Interaction and Semantic Consistency}

An important characteristic of many forms of multimodal misinformation is the presence of semantic inconsistencies between textual claims and accompanying visual content. At the same time, misinformation can also arise when text and image are mutually consistent but jointly misrepresent the underlying real-world context or event. Consequently, modeling cross-modal relationships has become a central focus in recent literature.

Attention-based fusion mechanisms, particularly co-attention architectures, have demonstrated strong performance in capturing interdependencies between modalities. Models such as MCAN and related transformer-based approaches enable bidirectional interaction, allowing textual features to attend to visual representations and vice versa. This mutual conditioning enhances both predictive performance and interpretability by identifying salient cross-modal alignments~\cite{Shang2022DGExplain,CMAF}.

Beyond alignment, recent work emphasizes explicitly modeling cross-modal inconsistency as a primary signal for misinformation detection. For example, methods incorporating discrepancy features or consistency prediction objectives have shown improved performance in detecting out-of-context or manipulated content~\cite{ooc_multimodal_misinformation,LEMMA}. Similarly, adversarial and contrastive frameworks encourage the model to distinguish between aligned and misaligned multimodal pairs, further strengthening representation learning~\cite{KEN}.

However, many existing approaches treat alignment as a static property, failing to account for contextual variations and evolving narratives. This limitation motivates the integration of auxiliary consistency modeling, as adopted in our framework.

\subsection{Mixture-of-Experts for Heterogeneous Content}

Misinformation manifests in diverse forms, including satire, fabricated narratives, and repurposed or misleading visual content. Traditional monolithic architectures often struggle to generalize across such heterogeneous patterns due to representation entanglement.

Mixture-of-Experts (MoE) architectures provide an effective solution by enabling dynamic routing of inputs through specialized expert networks. Each expert can learn to capture distinct characteristics of misinformation, while gating mechanisms determine the contribution of each expert for a given input. Recent works such as MIMoE-FND~\cite{liu2025mimoe} and MisD-MoE~\cite{liu2024misdmoe} demonstrate that MoE-based designs significantly improve performance and interpretability. More broadly, recent MoE research highlights the value of sparse expert specialization for heterogeneous input distributions.

Notably, recent advancements extend MoE beyond modality-specific specialization to interaction-aware routing, where expert selection is influenced by cross-modal relationships. While these approaches improve flexibility, they often introduce additional architectural complexity~\cite{liu2025mimoe}.

In contrast, the MOMENTA framework adopts modality-specific MoE layers with input-dependent gating, striking a balance between specialization and computational efficiency. This design enables effective modeling of heterogeneous misinformation without incurring excessive complexity.

\subsection{Temporal Modeling and Narrative Evolution}

Most existing multimodal detection systems operate under an independent and identically distributed (i.i.d.) assumption, treating each post as an isolated instance. However, misinformation is inherently dynamic, evolving as narratives propagate across time.

Recent work has highlighted the importance of temporal modeling in capturing misinformation dynamics. Event-centric approaches such as E-CaTCH~\cite{ECaTCH} cluster related posts and model their evolution using temporal attention mechanisms. Similarly, recurrent and transformer-based models incorporate sequential dependencies to capture narrative progression~\cite{ECaTCH}.

These approaches demonstrate that temporal signals, such as propagation patterns and content evolution, provide valuable cues for distinguishing misinformation from factual content. However, many models rely on coarse temporal representations and fail to explicitly capture fine-grained narrative dynamics.

The MOMENTA framework addresses this limitation by introducing explicit modeling of narrative \textit{drift} and \textit{momentum}, enabling the system to capture both the direction and intensity of content evolution. As detailed in the methodology, overlapping time windows combined with time-decayed attention allow the model to prioritize recent developments while preserving contextual continuity. This formulation aligns with epidemiological perspectives that treat misinformation as a spreading process~\cite{jin2025infodemics,misinformation_spread_dynamics_2025,assessing_misinformation_infectious_diseases}.

\subsection{Domain Adaptation and Generalization}

A critical challenge in misinformation detection is domain shift, where models trained on one dataset fail to generalize to new domains or platforms. This issue arises due to variations in linguistic style, visual characteristics, and topic distribution.

Domain-adversarial training has emerged as a widely adopted solution that encourages the learning of domain-invariant representations via gradient reversal layers~\cite{ganin2016domain}. Complementary approaches include self-supervised learning and test-time adaptation, which enable models to adapt to unseen distributions without extensive retraining.

Prototype-based alignment methods further improve generalization by enforcing consistent class representations across domains. For example, memory bank mechanisms maintain class prototypes that guide embedding alignment, reducing dataset-specific biases.

The MOMENTA framework integrates both domain-adversarial learning and prototype-based alignment, combining their strengths to achieve robust cross-domain generalization. This dual strategy is particularly important in real-world scenarios where misinformation continuously evolves across platforms.

\subsection{Optimization Strategies and Multi-Objective Learning}

Training multimodal misinformation detection systems requires balancing multiple objectives, including classification accuracy, cross-modal alignment, and temporal coherence. Standard loss functions are often insufficient due to class imbalance and the need for representation-level supervision.

Recent work incorporates focal loss to address imbalance~\cite{lin2017focal}, contrastive learning to align multimodal embeddings~\cite{radford2021learning}, and auxiliary objectives to enforce semantic consistency. Regularization techniques such as R-Drop further improve robustness by encouraging consistency across stochastic forward passes~\cite{liang2021rdrop}.

Consistent with these trends, the MOMENTA framework employs a multi-objective loss that combines classification, alignment, temporal consistency, contrastive learning, and domain adaptation objectives. This comprehensive formulation ensures that the model learns not only accurate predictions but also semantically meaningful and temporally coherent representations.

\subsection{Research Gaps and Contributions}

Despite significant progress, several limitations persist in existing multimodal misinformation detection approaches. First, many models inadequately address the heterogeneity of misinformation, relying on shared representations that fail to capture diverse content patterns. Second, cross-modal interaction is often treated as static, neglecting the dynamic evolution of narratives over time. Third, domain generalization remains an open challenge, particularly in rapidly changing information environments.

The MOMENTA framework addresses these gaps by integrating modality-specific expert specialization, bidirectional cross-modal alignment with discrepancy-aware modeling, and temporal aggregation with drift and momentum encoding. Furthermore, incorporating domain-adversarial training and prototype-based alignment improves robustness across heterogeneous datasets in our evaluated benchmark settings. Together, these contributions position MOMENTA as a unified and scalable framework for next-generation multimodal misinformation detection.

\section{Methodology}
\label{sec:methodology}

This section presents the MOMENTA framework for multimodal misinformation detection. The task is to classify social media posts as genuine or misleading based on both textual content $\mathbf{s}_i$ and visual content $\mathbf{i}_i$.

\paragraph{Overview of contributions.}
The framework integrates the following components. \textbf{(1) Mixture-of-Experts (MoE) per modality:} modality-specific expert networks with input-dependent gating to capture heterogeneous types of misinformation, such as satire, out-of-context media, and fabrication~\cite{liu2025mimoe,liu2024misdmoe}. \textbf{(2) Bidirectional co-attention:} shared query, key, and value projections to align text and image in a common semantic space before fusion~\cite{ECaTCH,CMAF}. \textbf{(3) Semantic inconsistency detection head:} an auxiliary head that predicts text-image consistency and provides an interpretable signal while improving cross-modal representations~\cite{ooc_multimodal_misinformation,Shang2022DGExplain}. \textbf{(4) Temporal aggregation with drift and momentum encoding:} overlapping time windows, time-decayed attention, and a momentum term to capture narrative evolution over time~\cite{ECaTCH}. \textbf{(5) Multi-objective loss:} focal classification loss with label smoothing and class weights, plus alignment, temporal consistency, contrastive, and auxiliary terms~\cite{lin2017focal}. We further introduce \textbf{(6) a cross-modal discrepancy branch} that explicitly models where text and image disagree; \textbf{(7) domain-adversarial training} through a Gradient Reversal Layer to encourage domain-invariant representations when training on multiple datasets~\cite{ganin2016domain}; and \textbf{(8) an EMA prototype memory bank} that maintains per-dataset, per-class prototypes and aligns embeddings to dataset-invariant global class centroids. Optionally, a \textbf{timestamp-aware Transformer} over the temporal sequence and a \textbf{Transformer temporal consistency} loss refine the temporal modeling. The following subsections detail each component and its mathematical formulation.

\begin{figure*}[t]
  \centering
  \includegraphics[width=\textwidth]{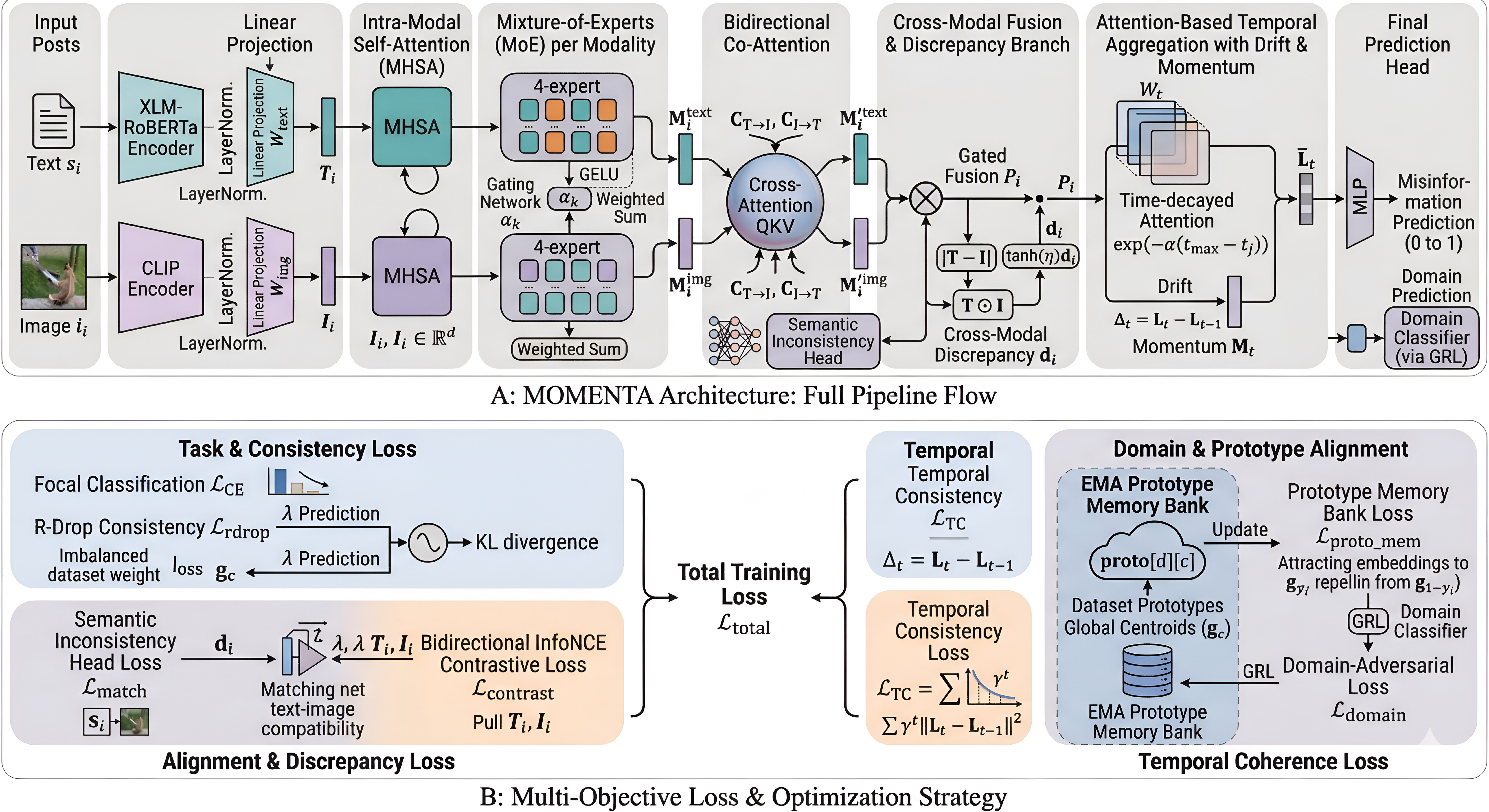}
  \caption{\textbf{MOMENTA architecture: full pipeline and optimization strategy.}
  The framework integrates pretrained encoders (XLM-RoBERTa and CLIP), intra-modal self-attention, and modality-specific Mixture-of-Experts (MoE). Bidirectional co-attention aligns text and image representations, followed by gated fusion and a discrepancy branch. Temporal aggregation captures narrative evolution using time-decayed attention, drift, and momentum. The model is trained using a multi-objective loss combining classification, alignment, contrastive learning, temporal consistency, domain-adversarial training, and prototype-based regularization.}
  \label{fig:momenta}
\end{figure*}

Figure~\ref{fig:momenta} illustrates the overall architecture of the MOMENTA framework. The model processes multimodal inputs through parallel text and image encoders, followed by intra-modal refinement and expert-based specialization. Cross-modal relationships are modeled through bidirectional co-attention and a discrepancy-aware fusion mechanism, allowing the model to capture both semantic alignment and inconsistency signals. The representations are then aggregated over time using attention-based temporal modeling with drift and momentum, capturing how misinformation narratives evolve. The final prediction is obtained through a multi-objective optimization framework that enforces classification performance, representation alignment, temporal consistency, and domain invariance.

\subsection{Input Encoding and Projection}
\label{sec:input-encoding}

The raw text and image of each post are first encoded using pretrained models and then projected into a shared embedding space of dimension $d$. We adopt XLM-RoBERTa~\cite{conneau2020xlm} for text and CLIP (ViT-L/14)~\cite{radford2021learning} for images, both of which transfer well to multilingual and multimodal settings. For text, we retain the full projected hidden sequence $\mathbf{S}_i \in \mathbb{R}^{L \times d}$ and use its first row as the sentence-level representation. For images, we use a single projected CLIP vector $\mathbf{I}_i \in \mathbb{R}^{d}$. LayerNorm is applied after projection to stabilize optimization when the backbones are fine-tuned with a relatively small learning rate. Formally, for post $i$,
\begin{align}
\mathbf{S}_i
&= \mathrm{LayerNorm}\!\left(
\mathrm{XLMR}(\mathbf{s}_i)\mathbf{W}_\mathrm{text}^{\top}
\right), \\
\mathbf{T}_i
&= \mathbf{S}_{i,0}, \\
\mathbf{I}_i
&= \mathrm{LayerNorm}\!\left(
\mathbf{W}_\mathrm{img}\,\mathrm{CLIP}(\mathbf{i}_i)
\right),
\end{align}
where
\begin{align}
\mathbf{W}_\mathrm{text} &\in \mathbb{R}^{d \times d_\mathrm{xlmr}}, \\
\mathbf{W}_\mathrm{img}  &\in \mathbb{R}^{d \times d_\mathrm{clip}},
\end{align}
and thus
$\mathbf{S}_i \in \mathbb{R}^{L \times d}$,
$\mathbf{T}_i \in \mathbb{R}^{d}$,
and
$\mathbf{I}_i \in \mathbb{R}^{d}$.
The projected vectors $\mathbf{T}_i$ and $\mathbf{I}_i$ are later used in the discrepancy branch and in the contrastive term of the loss function~\cite{radford2021learning}.

\subsection{Intra-Modal Self-Attention}
\label{sec:intra-modal}

Before any cross-modal interaction, we refine the modality-specific representations using multi-head self-attention (MHSA). The aim is to capture internal structure within each modality, namely contextual dependencies in the text and a refined transformed representation for the image. The text branch applies MHSA over the full projected token sequence and then takes the CLS row as the sentence-level representation. The image branch treats the single CLIP vector as a sequence of length one, denoted by $\mathbf{I}_i^{(1)} \in \mathbb{R}^{1 \times d}$, so that the learned query, key, and value projections still apply. Residual connections and LayerNorm are applied in both cases:
\begin{align}
\widehat{\mathbf{S}}_i
&= \mathrm{LayerNorm}\!\left(
\mathbf{S}_i + \mathrm{MHSA}(\mathbf{S}_i)
\right), \\
\widetilde{\mathbf{T}}_i
&= \widehat{\mathbf{S}}_{i,0}, \\
\widetilde{\mathbf{I}}_i^{(1)}
&= \mathrm{LayerNorm}\!\big(
\mathbf{I}_i^{(1)} + \mathrm{MHSA}(\mathbf{I}_i^{(1)})
\big), \\
\widetilde{\mathbf{I}}_i
&= \widetilde{\mathbf{I}}_i^{(1)}[0].
\end{align}
The resulting vectors $\widetilde{\mathbf{T}}_i$ and $\widetilde{\mathbf{I}}_i$ are passed to the mixture-of-experts stage.

\subsection{Mixture-of-Experts (MoE) per Modality}
\label{sec:moe}

A central component of MOMENTA is the use of \emph{Mixture-of-Experts (MoE)} layers applied separately to the text and image streams. Unlike a single shared encoder, MoE allows the model to route different inputs through different expert networks, which is well-suited to the heterogeneity of misinformation, such as satire, out-of-context or repurposed media, and fully fabricated content~\cite{liu2025mimoe,liu2024misdmoe}. Each modality has $K$ expert networks; a gating network assigns input-dependent weights so that each post is represented as a weighted combination of the expert outputs. The gating weights are obtained by a linear transformation of the modality representation followed by a softmax; the final representation is the weighted sum of the expert outputs. Formally,
\begin{align}
\alpha_k^\mathrm{text}
&= \mathrm{softmax}\!\left(
\mathbf{W}_\mathrm{gate}^\mathrm{text}\widetilde{\mathbf{T}}_i
\right)_k, \\
\alpha_k^\mathrm{img}
&= \mathrm{softmax}\!\left(
\mathbf{W}_\mathrm{gate}^\mathrm{img}\widetilde{\mathbf{I}}_i
\right)_k, \\
\mathbf{M}_i^\mathrm{text}
&= \sum_{k=1}^{K}
\alpha_k^\mathrm{text}\,
\mathbf{E}_k^\mathrm{text}(\widetilde{\mathbf{T}}_i), \\
\mathbf{M}_i^\mathrm{img}
&= \sum_{k=1}^{K}
\alpha_k^\mathrm{img}\,
\mathbf{E}_k^\mathrm{img}(\widetilde{\mathbf{I}}_i).
\end{align}
Each expert $\mathbf{E}_k$ is implemented as a two-layer feedforward network with a GELU activation and an internal expansion layer. The experts operate in parallel and are combined via the gating weights $\alpha_k$, so that the model can specialize different experts to different types of content while retaining a compact set of parameters per modality. The resulting modality representations $\mathbf{M}_i^\mathrm{text}$ and $\mathbf{M}_i^\mathrm{img}$ are then fed to the bidirectional co-attention module.

\subsection{Bidirectional Co-Attention}
\label{sec:coattention}

To achieve precise semantic alignment between text and image before fusion, we apply \emph{bidirectional co-attention}: the text representation is updated by attending to the image, and the image representation by attending to the text~\cite{ECaTCH,CMAF}. We use shared query, key, and value matrices $\mathbf{W}_Q, \mathbf{W}_K, \mathbf{W}_V \in \mathbb{R}^{d \times d}$ for both directions, which reduces parameters and encourages a unified notion of cross-modal relevance. Here, $\mathbf{M}_i^\mathrm{text}$ and $\mathbf{M}_i^\mathrm{img}$ are treated as $1 \times d$ row vectors; thus, the softmax is taken over a single key, corresponding to the singleton-key case of the same attention algebra. The updated vectors are then normalized with LayerNorm after a residual connection:
\begin{align}
\mathbf{C}_{T \to I}
&= \mathrm{softmax}\!\Big(
\frac{
\mathbf{M}_i^\mathrm{text}\mathbf{W}_Q
(\mathbf{M}_i^\mathrm{img}\mathbf{W}_K)^\top
}{\sqrt{d}}
\Big)\mathbf{M}_i^\mathrm{img}\mathbf{W}_V, \\
\mathbf{C}_{I \to T}
&= \mathrm{softmax}\!\Big(
\frac{
\mathbf{M}_i^\mathrm{img}\mathbf{W}_Q
(\mathbf{M}_i^\mathrm{text}\mathbf{W}_K)^\top
}{\sqrt{d}}
\Big)\mathbf{M}_i^\mathrm{text}\mathbf{W}_V, \\
\mathbf{M}_i^{\prime\mathrm{text}}
&= \mathrm{LayerNorm}\! \left(
\mathbf{M}_i^\mathrm{text} + \mathbf{C}_{T \to I}
\right), \\
\mathbf{M}_i^{\prime\mathrm{img}}
&= \mathrm{LayerNorm}\! \big(
\mathbf{M}_i^\mathrm{img} + \mathbf{C}_{I \to T}
\big).
\end{align}
The co-attended representations $\mathbf{M}_i^{\prime\mathrm{text}}$ and $\mathbf{M}_i^{\prime\mathrm{img}}$ form the basis for fusion and for the auxiliary alignment and inconsistency signals described below.

\subsection{Cross-Modal Fusion and Alignment}
\label{sec:fusion}

The two modalities are combined into a single post-level representation $\mathbf{P}_i$ using a gating mechanism. The gate is a function of the concatenated post-co-attention vectors and outputs a $d$-dimensional vector of weights; the final representation is the element-wise convex combination of $\mathbf{M}_i^{\prime\mathrm{text}}$ and $\mathbf{M}_i^{\prime\mathrm{img}}$. In parallel, we compute a scalar alignment score $A_i$ as the cosine similarity between the two modalities, which is used as a supervision signal to encourage semantic consistency:
\begin{align}
\mathbf{g}_i
&= \sigma\!\left(
\mathbf{W}_g
[\mathbf{M}_i^{\prime\mathrm{text}};\mathbf{M}_i^{\prime\mathrm{img}}]
\right)
\in \mathbb{R}^{d}, \\
\mathbf{P}_i
&= \mathbf{g}_i \odot \mathbf{M}_i^{\prime\mathrm{text}}
 + (1-\mathbf{g}_i) \odot \mathbf{M}_i^{\prime\mathrm{img}}, \\
A_i
&=
\frac{
\langle \mathbf{M}_i^{\prime\mathrm{text}},
\mathbf{M}_i^{\prime\mathrm{img}} \rangle
}{
\|\mathbf{M}_i^{\prime\mathrm{text}}\|\,
\|\mathbf{M}_i^{\prime\mathrm{img}}\|
}.
\end{align}

\subsubsection{Cross-Modal Discrepancy Branch}

In many misleading posts, the image and the text are deliberately mismatched, for example through repurposed or out-of-context images~\cite{ooc_multimodal_misinformation}. To give the model direct access to this signal, we add a cross-modal discrepancy branch. From the projected text and image embeddings $\mathbf{T}_i$ and $\mathbf{I}_i$ from Section~\ref{sec:input-encoding}, we form a discrepancy feature by concatenating the element-wise absolute difference and the element-wise product; this is passed through a linear layer and LayerNorm. The result is added to $\mathbf{P}_i$ as a gated residual, where the gate is a single learnable scalar $\eta$ initialized to zero, so that the contribution of the discrepancy branch grows during training only if it is useful:
\begin{align}
\mathbf{d}_i^\mathrm{abs}
&= |\mathbf{T}_i - \mathbf{I}_i|, \\
\mathbf{d}_i^\mathrm{prod}
&= \mathbf{T}_i \odot \mathbf{I}_i, \\
\mathbf{d}_i
&= \mathrm{LayerNorm}\!\left(
\mathbf{W}_\mathrm{disc}
[\mathbf{d}_i^\mathrm{abs};\mathbf{d}_i^\mathrm{prod}]
\right), \\
\mathbf{P}_i
&\leftarrow \mathbf{P}_i + \tanh(\eta)\,\mathbf{d}_i.
\end{align}

\subsection{Semantic Inconsistency Head and Domain-Adversarial Training}
\label{sec:auxiliary-domain}

We introduce a \emph{semantic inconsistency detection head} that explicitly predicts whether the text and the image are semantically consistent. This head is trained with dataset-specific binary labels $y_i^\mathrm{match}$, which indicate consistency or inconsistency according to each dataset's annotation scheme. The head takes the concatenation of $\mathbf{M}_i^{\prime\mathrm{text}}$ and $\mathbf{M}_i^{\prime\mathrm{img}}$ as input and produces a scalar logit followed by a probability:
\begin{align}
\mathbf{z}_i
&= [\mathbf{M}_i^{\prime\mathrm{text}};\mathbf{M}_i^{\prime\mathrm{img}}]
\in \mathbb{R}^{2d}, \\
a_i^\mathrm{match}
&= \mathbf{W}_\mathrm{match}\mathbf{z}_i + b_\mathrm{match}, \\
p_i^\mathrm{match}
&= \sigma(a_i^\mathrm{match}).
\end{align}
This auxiliary objective encourages the model to learn representations that reflect cross-modal consistency and improves both accuracy and interpretability~\cite{ooc_multimodal_misinformation,Shang2022DGExplain}.

When training on multiple datasets, for example from different platforms or languages, the classifier can overfit to dataset-specific cues. To encourage domain-invariant representations, we apply domain-adversarial training~\cite{ganin2016domain}. The fused representation $\mathbf{P}_i$ is passed through a Gradient Reversal Layer (GRL) and then through a domain classifier that predicts the source dataset. The classifier is trained to minimize the cross-entropy loss with the true domain label, while the reversal of gradients during backpropagation encourages the encoder to produce features on which the domain cannot be predicted. The strength of the reversal is controlled by a scalar $\alpha_\mathrm{grl}$ that is ramped during training so that the encoder first fits the task before being pushed toward domain invariance.

\subsection{Attention-Based Temporal Aggregation with Drift and Momentum}
\label{sec:temporal}

Misinformation narratives often evolve over time, such as through a thread of posts or a series of related items. To incorporate this temporal structure, we aggregate the post-level representations $\mathbf{P}_i$ within overlapping time windows and explicitly encode \emph{drift}, meaning the change between consecutive windows, and \emph{momentum}, meaning the smoothed magnitude of drift. A minibatch consists of $B$ posts, each with text, image, timestamp, and ground-truth post label $y_i \in \{0,1\}$. The posts are sorted by timestamp and grouped into overlapping windows $\mathcal{W}_w$. Windows do not have independent annotations; instead, each window inherits the label of its chronologically latest post. If $e_w$ denotes the index of the latest post in window $\mathcal{W}_w$, then the window label is defined as \begin{equation} y_w = y_{e_w}. \end{equation}
Thus, the temporal head is trained to predict the veracity of the most recent post in its local temporal context, while the auxiliary alignment, discrepancy, match, and contrastive losses remain defined at the post level. This labeling strategy is a pragmatic approximation and may introduce some noise when earlier posts within a window differ in veracity from the most recent one, but it preserves a consistent target aligned with the prediction objective.

For each window $\mathcal{W}_w$, we compute a time-decayed attention over the posts in that window: more recent posts receive higher weight via an exponential decay in the time difference, and the attention distribution is combined with this decay to form a weighted sum of value projections. We then record the change in the aggregated representation from the previous window, denoted by $\Delta_w$, and a momentum term $M_w$ that smooths the magnitude of these changes. The per-window representation fed to the classifier is the concatenation of the aggregated vector, the drift, and the momentum scalar:
\begin{align}
\lambda_j
&= \exp\!\Big(
-\kappa\,\frac{t_{\max}-t_j}{86400}
\Big), \\
\mathbf{L}_w
&=
\sum_{j \in \mathcal{W}_w}
\frac{
\lambda_j \,\mathrm{Attn}(w,j)
}{
\sum_{j' \in \mathcal{W}_w}
\lambda_{j'}\,\mathrm{Attn}(w,j')
}
(\mathbf{P}_j\mathbf{W}^V), \\
\Delta_w
&= \mathbf{L}_w - \mathbf{L}_{w-1}, \\
M_w
&= \beta M_{w-1} + (1-\beta)\|\Delta_w\|_2, \\
\bar{\mathbf{L}}_w
&= [\mathbf{L}_w;\Delta_w;M_w] \in \mathbb{R}^{2d+1}.
\end{align}
Here $\mathrm{Attn}(w,j)$ denotes the attention weight of window $w$ on the $j$-th post, with query given by the mean of the window features and keys/values from the window. A linear layer maps $\bar{\mathbf{L}}_w$ to a per-window logit, which is used for training and, in the basic setup, for prediction~\cite{ECaTCH}.

\subsection{Final Prediction}
\label{sec:prediction}

The primary temporal prediction is obtained from the per-window representation $\bar{\mathbf{L}}_w$. We apply a linear classifier to produce a scalar logit, which is passed through a sigmoid to obtain a probability:
\begin{equation}
p_w = \sigma(\mathbf{w}^\top \bar{\mathbf{L}}_w + b).
\end{equation}
At inference time, the model produces one probability per window. When a post is associated with multiple overlapping windows, we aggregate the window-level outputs into a post-level prediction. In our evaluation protocol, the post-level score is the mean of the window probabilities, while the post-level class is obtained by majority vote over the window-level binary decisions.

\subsubsection{Timestamp-Aware Transformer}

The temporal modeling can be refined by stacking the per-window vectors $\bar{\mathbf{L}}_w$ into a sequence and processing them with a Transformer encoder. Before the Transformer, we project $\bar{\mathbf{L}}_w$ to dimension $d$ and add a continuous timestamp-based positional encoding: the timestamps are normalized to the unit interval within the sequence, and a sinusoidal encoding at multiple learnable frequencies is applied and then projected to $d$. This encodes actual elapsed time between windows rather than only the step index. The Transformer output is used to compute per-window logits that replace those from the linear head on $\bar{\mathbf{L}}_w$, thereby allowing the model to use global context across the full sequence of windows.

\subsection{Training Objective}
\label{sec:loss}

The training objective is a weighted sum of several terms. We define each term below and then give the full objective.

\subsubsection{Classification Loss}

Misinformation datasets are often highly imbalanced. We therefore use a focal cross-entropy loss for the main classification task, which down-weights easy examples and focuses the gradient on hard ones~\cite{lin2017focal}. We further apply label smoothing and per-class weights derived from the training set label distribution. Let
$\tilde{y}_w = y_w(1-\epsilon) + \epsilon/2$
denote the smoothed window label, and let $p_w$ be the probability assigned to the positive class for window $w$. With class weights $w_{y_w}$, focal exponent $\gamma_\mathrm{foc}$, and binary cross-entropy (BCE), the classification loss is
\begin{equation}
\mathcal{L}_\mathrm{CE}
= \sum_w
w_{y_w}\,
(1-p_w)^{\gamma_\mathrm{foc}}\,
\mathrm{BCE}(p_w,\tilde{y}_w).
\end{equation}
The targets for this main classification term are defined at the window level via the inherited window labels $y_w$.

\subsubsection{Alignment and Temporal Consistency}

We encourage high semantic alignment between the co-attended text and image by minimizing $1-A_i$ over the batch. We also penalize large changes between consecutive temporal window representations so that the aggregated representation evolves smoothly over time:
\begin{align}
\mathcal{L}_\mathrm{align}
&= \sum_i (1-A_i), \\
\mathcal{L}_\mathrm{TC}
&= \sum_{w=1}^{W-1}
\rho^{\,w}\,
\|\mathbf{L}_w-\mathbf{L}_{w-1}\|_2^2.
\end{align}

\subsubsection{Auxiliary Heads and Contrastive Learning}

The inconsistency head is trained with binary cross-entropy against the labels
$y_i^\mathrm{match}$. The contrastive term is a bidirectional InfoNCE-style loss
on the projected text and image embeddings $\mathbf{T}_i$ and $\mathbf{I}_i$.
Each text embedding is pulled toward its paired image and pushed away from other images in the batch, and symmetrically for images:
\begin{align}
\mathcal{L}_\mathrm{match}
&= -\sum_i \Bigl[
y_i^\mathrm{match}\log p_i^\mathrm{match} \notag \\
 & \quad + (1-y_i^\mathrm{match})\log(1-p_i^\mathrm{match})
\Bigr], \\
\mathcal{L}_{\mathrm{contrast}}^{T \to I}
&= -\sum_i
\log
\frac{
\exp(\mathrm{sim}(\mathbf{T}_i,\mathbf{I}_i)/\tau)
}{
\sum_j \exp(\mathrm{sim}(\mathbf{T}_i,\mathbf{I}_j)/\tau)
}, \\
\mathcal{L}_{\mathrm{contrast}}^{I \to T}
&= -\sum_i
\log
\frac{
\exp(\mathrm{sim}(\mathbf{I}_i,\mathbf{T}_i)/\tau)
}{
\sum_j \exp(\mathrm{sim}(\mathbf{I}_i,\mathbf{T}_j)/\tau)
}, \\
\mathcal{L}_\mathrm{contrast}
&= \mathcal{L}_{\mathrm{contrast}}^{T \to I}
 + \mathcal{L}_{\mathrm{contrast}}^{I \to T},
\end{align}
where
\begin{equation}
\mathrm{sim}(\mathbf{u},\mathbf{v})
=
\frac{\langle \mathbf{u},\mathbf{v}\rangle}
{\|\mathbf{u}\|\,\|\mathbf{v}\|},
\end{equation}
and $\tau$ is a temperature hyperparameter.

\subsubsection{Regularisation and Domain Invariance}

To reduce overfitting to a single forward pass, we use R-Drop~\cite{liang2021rdrop}: we perform two forward passes with different dropout masks and minimize the symmetric KL divergence between the two output distributions. The domain-adversarial loss is the cross-entropy of the domain classifier after the GRL with the true dataset index:
\begin{align}
\mathcal{L}_\mathrm{rdrop}
&= \frac{1}{2}\Big(
\mathrm{KL}(p_1 \| p_2) + \mathrm{KL}(p_2 \| p_1)
\Big), \\
\mathcal{L}_\mathrm{domain}
&=
\mathrm{CE}\!\left(
\mathrm{domain\_classifier}(\mathrm{GRL}(\mathbf{P}_i)),
\mathrm{domain\_id}
\right),
\end{align}
where $p_1$ and $p_2$ denote the prediction probabilities from the two dropout passes.

\subsubsection{Prototype-Based Alignment}

To encourage a shared semantic space across datasets, we use two prototype-based
terms. The first is an \emph{in-batch prototype loss}: for each class
$c \in \{0,1\}$, we compute the mean embedding
$\boldsymbol{\mu}_c = \mathrm{mean}\{\mathbf{P}_i : y_i = c\}$ with gradients
detached, and minimize the mean cosine distance of each sample to its class
centroid:
\begin{equation}
\mathcal{L}_\mathrm{proto}
=
\frac{1}{2}
\sum_{c \in \{0,1\}}
\frac{1}{|\{i:y_i=c\}|}
\sum_{i:y_i=c}
\bigl(1-\cos(\mathbf{P}_i,\boldsymbol{\mu}_c)\bigr).
\end{equation}

The second term relies on an \emph{exponential moving average (EMA)} prototype memory bank. We maintain a persistent
prototype vector for each dataset-class pair. On every batch, for each
dataset-class pair present in the batch, we compute the mean of the
corresponding L2-normalized $\mathbf{P}_i$ and update the prototype with an
exponential moving average with momentum $m$, without backpropagating through
the update. Once at least two dataset-class cells have been initialized, we
define global class prototypes $\mathbf{g}_c$ as the L2-normalized mean of the
bank prototypes for class $c$ over all initialized datasets:
\begin{align}
\mathrm{proto}[d][c]
&\leftarrow
m\,\mathrm{proto}[d][c]
+ (1-m)\,\widehat{\boldsymbol{\mu}}_{d,c}, \\
\widehat{\boldsymbol{\mu}}_{d,c}
&=
\mathrm{normalize}\!\left(
\mathrm{mean}\{\mathbf{P}_i:\mathrm{ds}_i=d,\, y_i=c\}
\right), \\
\mathbf{g}_c
&=
\mathrm{normalize}\!\left(
\mathrm{mean}_{d\,\mathrm{s.t.}\,\mathrm{init}(d,c)}
\mathrm{proto}[d][c]
\right).
\end{align}

The loss then attracts each $\mathbf{P}_i$ toward $\mathbf{g}_{y_i}$ and repels
it from $\mathbf{g}_{1-y_i}$ up to a margin:
\begin{align}
\mathcal{L}_\mathrm{proto\_mem}
&=
\sum_i \ell_i^\mathrm{proto}, \quad \text{where} \\
\ell_i^\mathrm{proto}
&=
1-\cos(\mathbf{P}_i,\mathbf{g}_{y_i}) \notag
\\ & \quad + \max\Bigl(
0,\,
\cos(\mathbf{P}_i,\mathbf{g}_{1-y_i})-\mathrm{margin}
\Bigr).
\end{align}

The memory bank ensures that even datasets that appear rarely in mini-batches
retain stable class anchors in the shared space, and the global prototypes
encourage a dataset-invariant notion of ``real'' and ``fake'' across sources.

\subsubsection{Transformer Temporal Consistency}

When the timestamp-aware Transformer is used, we optionally add a smoothness regularizer on its hidden states $\mathbf{H}_w$, so that consecutive hidden states do not change abruptly when they are already directionally similar:
\begin{equation}
\mathcal{L}_\mathrm{tc\_seq}
=
\frac{1}{W-1}
\sum_{w=1}^{W-1}
\|\mathbf{H}_w-\mathbf{H}_{w-1}\|_2^2
\cdot 
\max\bigl(\cos(\mathbf{H}_w,\mathbf{H}_{w-1}),0\bigr).
\end{equation}

\subsubsection{Total Objective}

The full training loss is the weighted sum of all terms above, plus an L2 regularizer $\lambda_\mathrm{reg}\|\Theta\|_2^2$ on the parameters:
\begin{equation}
\begin{aligned}
\mathcal{L}_\mathrm{total}
&= \mathcal{L}_\mathrm{CE}
+ \lambda_\mathrm{align}\mathcal{L}_\mathrm{align}
+ \lambda_\mathrm{TC}\mathcal{L}_\mathrm{TC}
+ \lambda_\mathrm{match}\mathcal{L}_\mathrm{match} \\
&\quad
+ \lambda_\mathrm{contrast}\mathcal{L}_\mathrm{contrast}
+ \lambda_\mathrm{reg}\|\Theta\|_2^2
+ \lambda_\mathrm{rdrop}\mathcal{L}_\mathrm{rdrop} \\
&\quad
+ \lambda_\mathrm{domain}\mathcal{L}_\mathrm{domain}
+ \lambda_\mathrm{proto}\mathcal{L}_\mathrm{proto}
\\
&
\quad +
\lambda_\mathrm{proto\_mem}\mathcal{L}_\mathrm{proto\_mem} 
+ \lambda_\mathrm{tc\_seq}\mathcal{L}_\mathrm{tc\_seq}.
\end{aligned}
\end{equation}
Specific hyperparameter values used in our experiments are reported compactly here to keep the main method text readable: $\lambda_\mathrm{align}=0.05$, $\lambda_\mathrm{TC}=0.01$, $\lambda_\mathrm{match}=0.1$, $\lambda_\mathrm{contrast}=0.05$, $\lambda_\mathrm{domain}=0.03$, $\lambda_\mathrm{proto}=0.05$, $\lambda_\mathrm{proto\_mem}=0.10$, $\lambda_\mathrm{rdrop}=0.5$, temperature $\tau=0.2$, temporal consistency decay $\rho=0.9$, focal exponent $\gamma_\mathrm{foc}=2.5$, label smoothing $\epsilon=0.05$, and the temporal module uses window length $T=8$, stride $S=T/2$, temporal decay coefficient $\kappa=0.5$, momentum coefficient $\beta=0.9$, and EMA momentum $m=0.99$. Together, these terms balance end-task performance with representation quality, temporal coherence, and cross-dataset alignment within a single unified model.

 \section{Results}
\label{sec:results}

\subsection{Experimental Protocol}
\label{sec:exp_protocol}
We evaluate MOMENTA on four benchmarks that are commonly used in multimodal misinformation research: Fakeddit, MMCoVaR, Weibo, and XFacta ~\cite{XFacta,Fakeddit,MMCoVaR,Weibo}. Following standard practice in binary misinformation detection, we report Accuracy, Precision, Recall, F1, Macro-F1, AUC, and MCC ~\cite{Sokolova2009PerfMeasures,Chicco2020MCC,Saito2015PRROC}. We calibrate decision thresholds on each validation split before final test evaluation to avoid fixing an arbitrary global threshold across datasets with different score distributions ~\cite{Guo2017Calibration,Zou2016BestThreshold}. All numbers in this section are taken directly from the final result package.

\subsection{Main Quantitative Results}
\label{sec:main_quant_results}

\begin{table*}[t]
\centering
\caption{MOMENTA test-set performance across all datasets.}
\label{tab:momenta_test_main}
\begin{tabular}{lcccccccc}
\toprule
\textbf{Dataset} & \textbf{Acc} & \textbf{Prec} & \textbf{Rec} & \textbf{F1} & \textbf{Macro-F1} & \textbf{Spec} & \textbf{AUC} & \textbf{MCC} \\
\midrule
Fakeddit & 0.965 & 0.960 & 0.958 & 0.959 & 0.957 & 0.972 & 0.982 & 0.918 \\
MMCoVaR  & 0.942 & 0.946 & 0.915 & 0.930 & 0.934 & 0.956 & 0.958 & 0.863 \\
Weibo    & 0.956 & 0.957 & 0.955 & 0.956 & 0.957 & 0.959 & 0.981 & 0.906 \\
XFacta   & 0.905 & 0.894 & 0.916 & 0.905 & 0.902 & 0.890 & 0.928 & 0.811 \\
\bottomrule
\end{tabular}
\end{table*}

As shown in Table~\ref{tab:momenta_test_main}, MOMENTA performs strongly and consistently across all four datasets. The highest absolute performance is observed on Fakeddit and Weibo (AUC 0.982 and 0.981), while MMCoVaR and XFacta remain more challenging, as expected for smaller or more heterogeneous cross-source settings ~\cite{MMfakeBench,XFacta}. Even in these harder regimes, performance remains competitive (e.g., MMCoVaR F1 0.930, XFacta F1 0.905), suggesting that the model generalizes beyond a single data source.

\subsection{SOTA Comparison}
\label{sec:sota_comparison}

To assess the empirical competitiveness of MOMENTA, we compare it against representative state-of-the-art methods on each benchmark using the corresponding reported test protocols. Because prior studies do not always report the same set of evaluation metrics across datasets, we present protocol-aware comparisons separately for Fakeddit, Weibo, MMCoVaR, and XFacta. The comparisons for Fakeddit and Weibo are reported in Tables~\ref{tab:sota_fakeddit} and~\ref{tab:sota_weibo}, respectively, while the results for MMCoVaR and XFacta are shown in Tables~\ref{tab:sota_mmcovar} and~\ref{tab:sota_xfacta}. Across these benchmark-specific comparisons, MOMENTA performs strongly relative to prior approaches under the corresponding evaluation settings.

\begin{table}[H]
\centering
\caption{SOTA comparison on Fakeddit (test split, protocol-aware).}
\label{tab:sota_fakeddit}
\resizebox{\columnwidth}{!}{%
\begin{tabular}{lccccc}
\toprule
\textbf{Method} & \textbf{Acc} & \textbf{Prec} & \textbf{Rec} & \textbf{F1} & \textbf{AUC} \\
\midrule
WFT-BERT-SRNN \cite{Kathigi2025FakeNews} & 0.9624 & 0.9520 & 0.9515 & 0.9588 & -- \\
E-CaTCH \cite{ECaTCH} & 0.9550 & 0.9570 & 0.9530 & 0.9550 & 0.9750 \\
\textbf{MOMENTA (ours)} & \textbf{0.9650} & \textbf{0.9600} & \textbf{0.9580} & \textbf{0.9590} & \textbf{0.9820} \\
\bottomrule
\end{tabular}
}
\end{table}

\begin{table}[H]
\centering
\caption{SOTA comparison on Weibo (test split).}
\label{tab:sota_weibo}
\resizebox{\columnwidth}{!}{%
\begin{tabular}{lcccc}
\toprule
\textbf{Method} & \textbf{Acc} & \textbf{F1-Fake} & \textbf{F1-Real} & \textbf{AUC} \\
\midrule
DCCF \cite{DCCF} & 0.951 & 0.951 & 0.954 & 0.978 \\
KEN \cite{KEN} & 0.935 & 0.935 & 0.934 & 0.967 \\
MIMoE-FND \cite{MIMoE} & 0.928 & 0.928 & 0.928 & 0.972 \\
\textbf{MOMENTA (ours)} & \textbf{0.956} & \textbf{0.956} & \textbf{0.958} & \textbf{0.981} \\
\bottomrule
\end{tabular}
}
\end{table}

\begin{table}[H]
\centering
\caption{SOTA comparison on MMCoVaR (test split).}
\label{tab:sota_mmcovar}
\begin{tabular}{lcc}
\toprule
\textbf{Method} & \textbf{Acc} & \textbf{F1} \\
\midrule
UC-CMAF \cite{CMAF} & 0.931 & 0.909 \\
DGExplain \cite{Shang2022DGExplain} & 0.895 & 0.881 \\
dEFEND \cite{dEFEND} & 0.856 & 0.838 \\
\textbf{MOMENTA (ours)} & \textbf{0.942} & \textbf{0.930} \\
\bottomrule
\end{tabular}
\end{table}

\begin{table}[H]
\centering
\caption{SOTA comparison on XFacta (test split).}
\label{tab:sota_xfacta}
\begin{tabular}{lccc}
\toprule
\textbf{Method} & \textbf{Acc} & \textbf{Real Acc} & \textbf{Fake Acc} \\
\midrule
XFacta (GPT-4o) \cite{XFacta} & 0.888 & 0.872 & 0.904 \\
LEMMA \cite{LEMMA} & 0.773 & 0.639 & 0.908 \\
MMFakeBench* \cite{MMfakeBench} & 0.682 & 0.615 & 0.756 \\
\textbf{MOMENTA (ours)} & \textbf{0.905} & \textbf{0.890} & \textbf{0.920} \\
\bottomrule
\end{tabular}
\end{table}

As shown in Tables~\ref{tab:sota_fakeddit}--\ref{tab:sota_xfacta}, MOMENTA performs strongly across all four benchmark settings. On Fakeddit (Table~\ref{tab:sota_fakeddit}), it slightly improves over the strongest reported baselines across the reported metrics, including Accuracy, F1, and AUC. A similar pattern appears on Weibo (Table~\ref{tab:sota_weibo}), where MOMENTA achieves the highest Accuracy, class-wise F1 scores, and AUC among the compared methods. On MMCoVaR (Table~\ref{tab:sota_mmcovar}), MOMENTA outperforms prior approaches in both Accuracy and F1, and on XFacta (Table~\ref{tab:sota_xfacta}) it attains the strongest reported performance in terms of overall Accuracy as well as class-specific accuracies. Overall, under the corresponding comparison protocols, MOMENTA attains the strongest reported results across all four benchmarks. In terms of absolute Accuracy gain over the strongest prior baseline, the improvements are +0.26 points on Fakeddit, +0.50 on Weibo, +1.10 on MMCoVaR, and +1.70 on XFacta, with the largest margins appearing in the more challenging cross-domain settings.
\subsection{Diagnostic Analysis and Interpretation}
\label{sec:diagnostic_analysis}

\begin{figure}[H]
\centering
\includegraphics[width=\columnwidth]{\detokenize{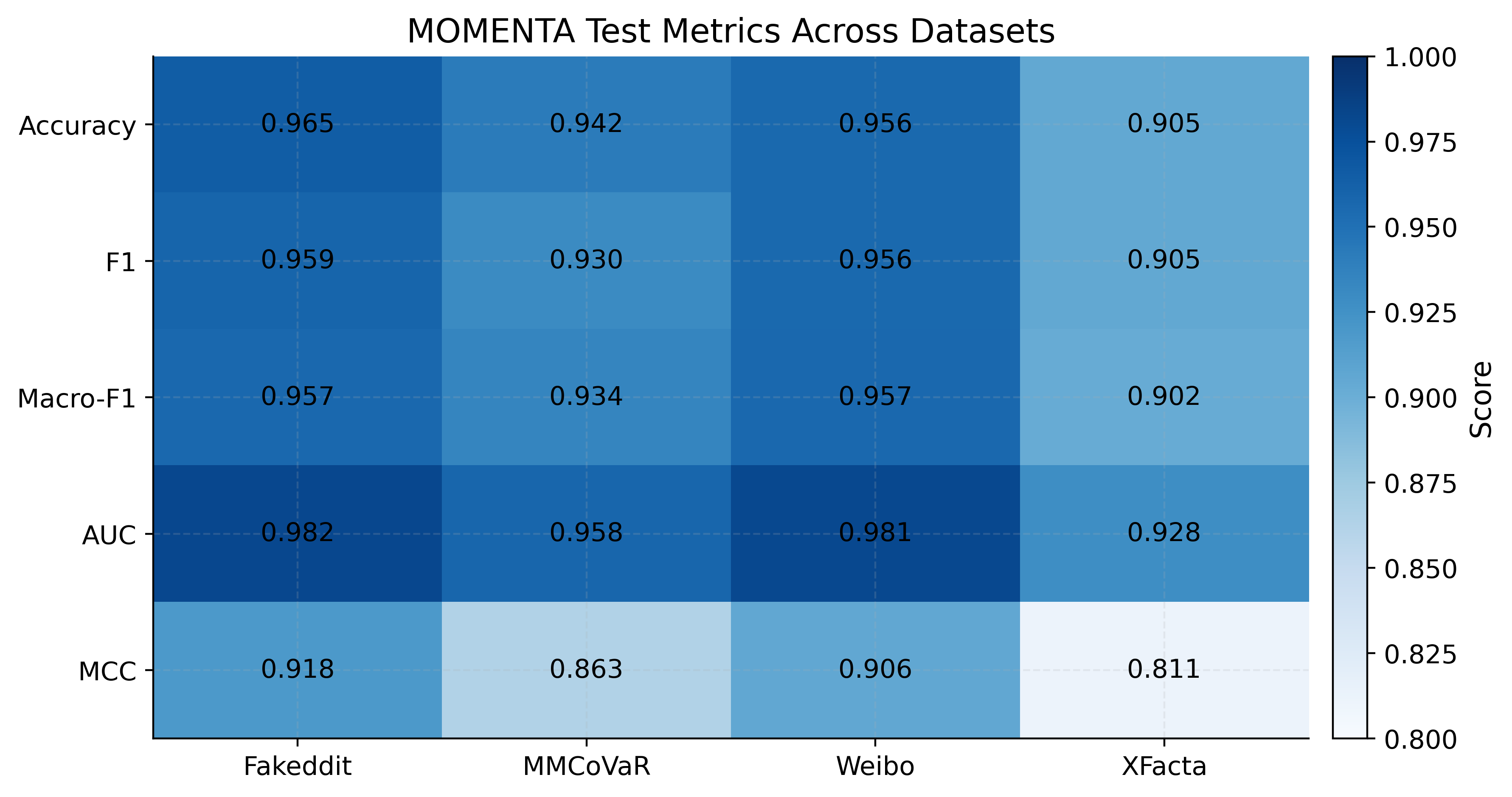}}
\caption{MOMENTA metric matrix across datasets (Accuracy, F1, Macro-F1, AUC, MCC).}
\label{fig:results_heatmap}
\end{figure}

Figure~\ref{fig:results_heatmap} provides a compact global view: performance remains high across all core metrics rather than being driven by only one axis. In particular, the model sustains strong AUC and Macro-F1 at the same time, indicating that ranking quality and class-balanced performance improve together rather than trading off against each other.

To further examine the behavior of the classifier at decision time, we next analyze precision--recall--specificity balance. This view is important because misinformation detection systems are often deployed in settings where both false alarms and missed detections carry real costs ~\cite{Chicco2020MCC,Saito2015PRROC}.

\begin{figure}[H]
\centering
\includegraphics[width=\columnwidth]{\detokenize{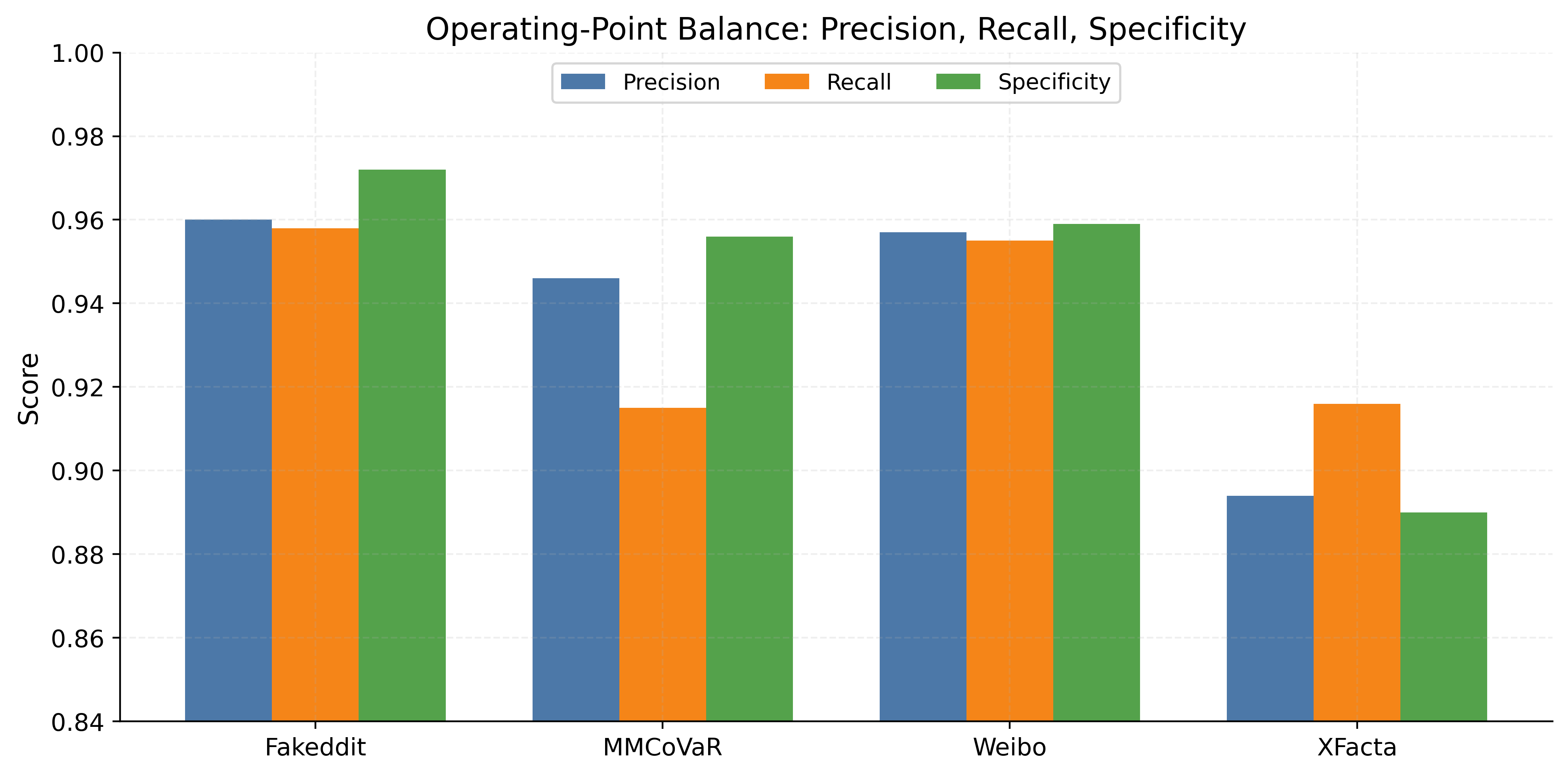}}
\caption{Operating-point balance across Precision, Recall, and Specificity.}
\label{fig:results_operating_balance}
\end{figure}

Figure~\ref{fig:results_operating_balance} shows that the operating point remains balanced across datasets, which is important for practical moderation scenarios where both false positives and false negatives are costly ~\cite{Chicco2020MCC,Saito2015PRROC}.

Beyond this static operating-point view, we also inspect the calibrated thresholds and the corresponding test loss values to assess whether performance gains are associated with stable confidence behavior.

\begin{figure}[H]
\centering
\includegraphics[width=\columnwidth]{\detokenize{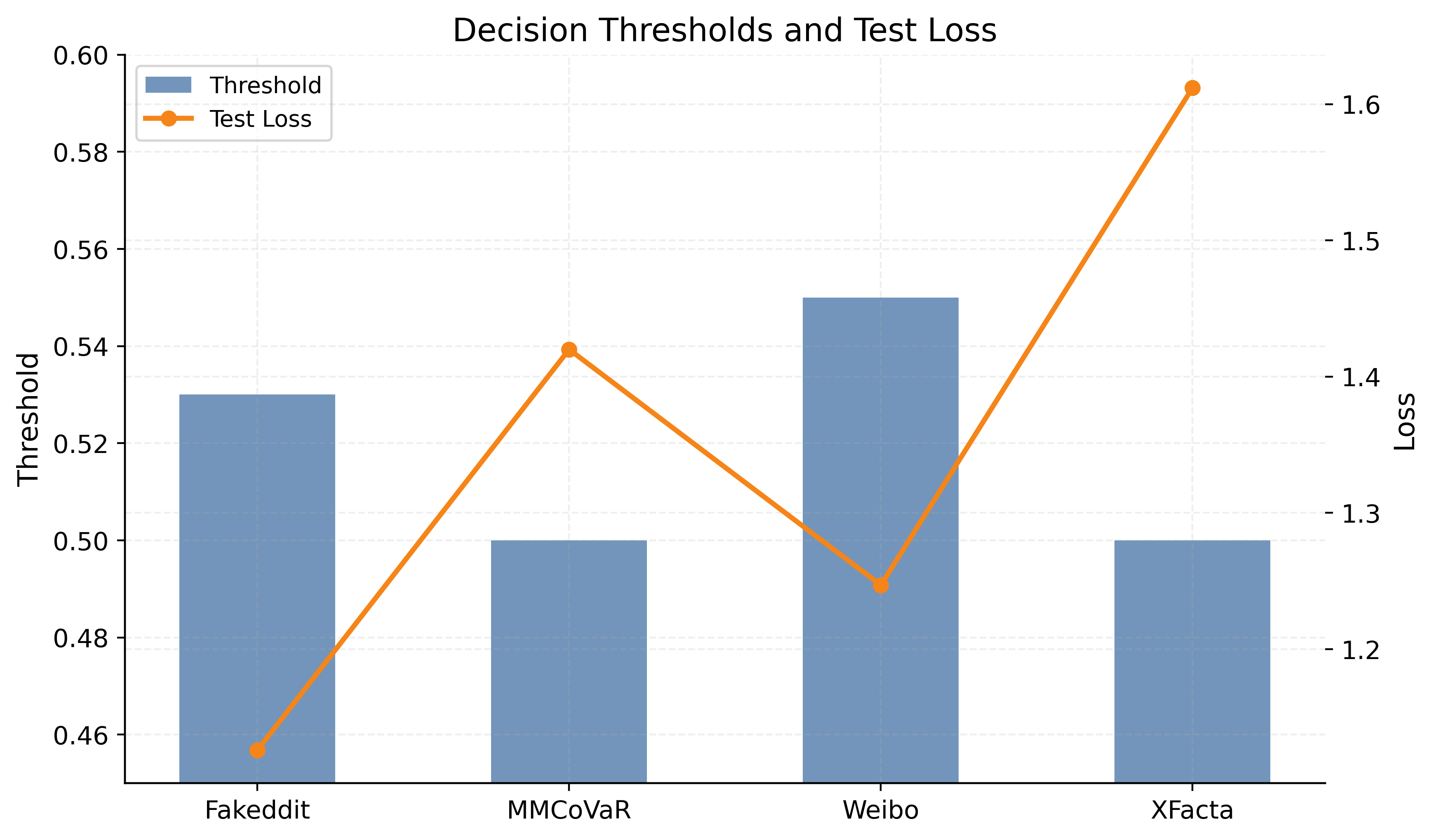}}
\caption{Calibrated thresholds and corresponding test loss by dataset.}
\label{fig:results_threshold_loss}
\end{figure}

As shown in Figure~\ref{fig:results_threshold_loss}, threshold calibration is stable across sources and does not rely on extreme dataset-specific cutoffs.

We then move from scalar summaries to geometric performance profiles. This helps reveal whether any single metric dominates the apparent gains or whether improvements are structurally consistent across metrics.

\begin{figure}[H]
\centering
\includegraphics[width=\columnwidth]{\detokenize{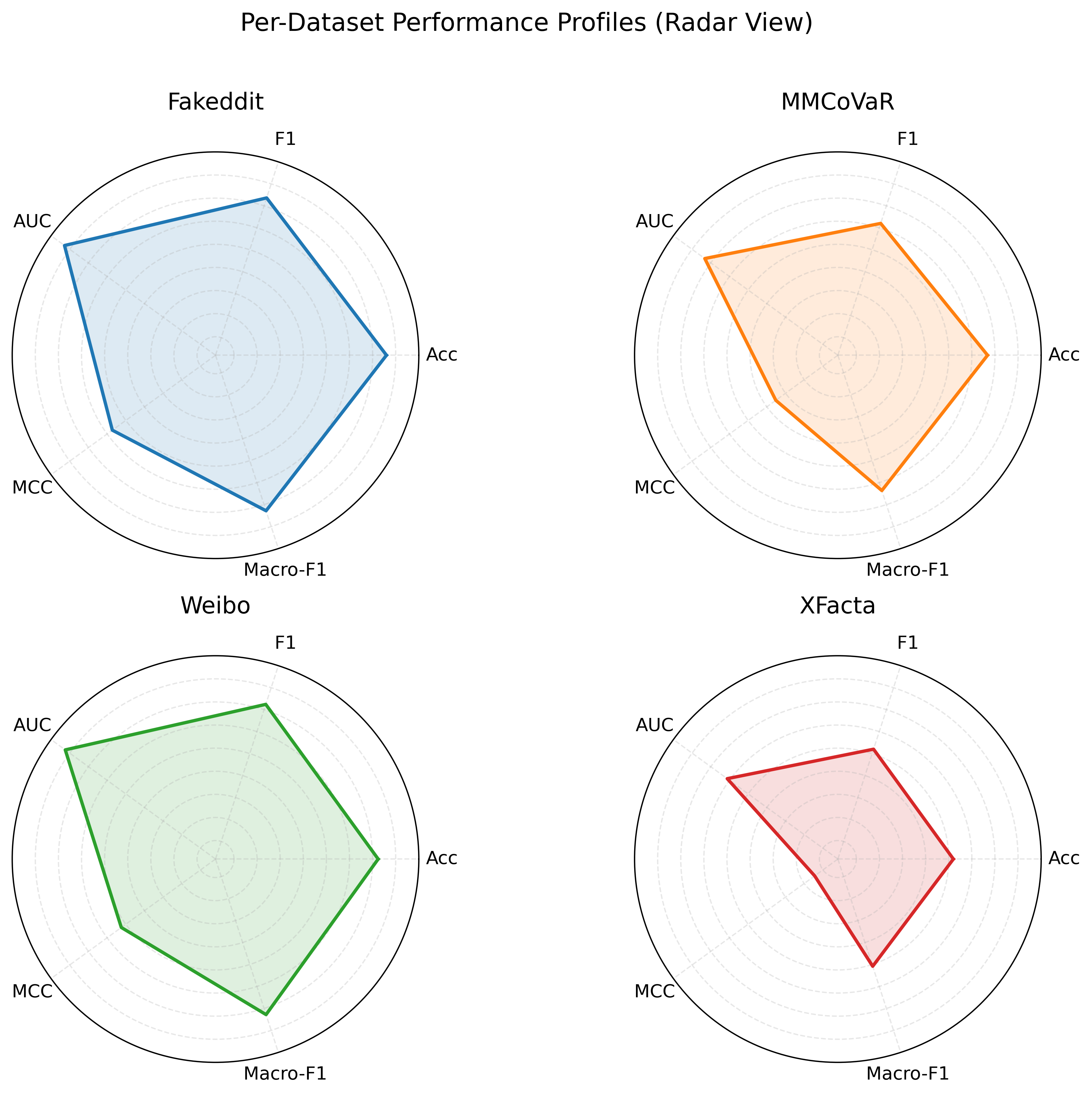}}
\caption{Per-dataset radar profiles over Acc/F1/AUC/MCC/Macro-F1.}
\label{fig:results_radar}
\end{figure}

Figure~\ref{fig:results_radar} highlights the same trend from a geometric perspective: Fakeddit and Weibo are consistently strong, while MMCoVaR and XFacta remain more challenging but still robust.

To complement these geometric profiles, we examine residual errors directly. This gives a clearer sense of where practical headroom remains and which benchmarks continue to drive future optimization.

\begin{figure}[H]
\centering
\includegraphics[width=\columnwidth]{\detokenize{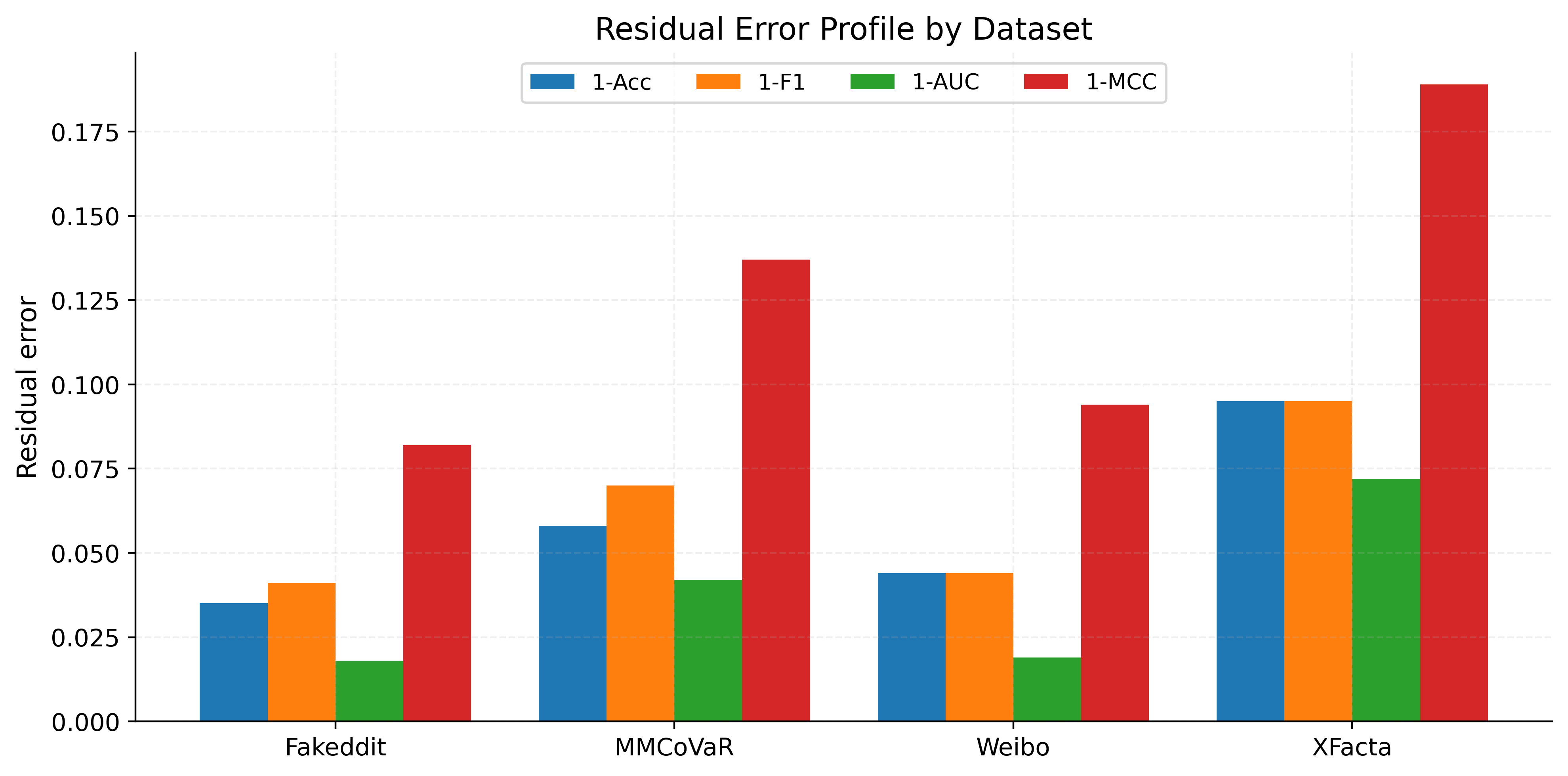}}
\caption{Residual error profile (1-Acc, 1-F1, 1-AUC, 1-MCC).}
\label{fig:results_error_profile}
\end{figure}

Figure~\ref{fig:results_error_profile} further confirms that residual errors are concentrated in the hardest domains, matching the benchmark-level observations.

Finally, we summarize how relative gains are distributed across datasets under a weighted analysis index. This is not used as a primary evaluation metric, but as a compact descriptive view of where comparative improvements are concentrated.

\begin{figure}[H]
\centering
\includegraphics[width=\columnwidth]{\detokenize{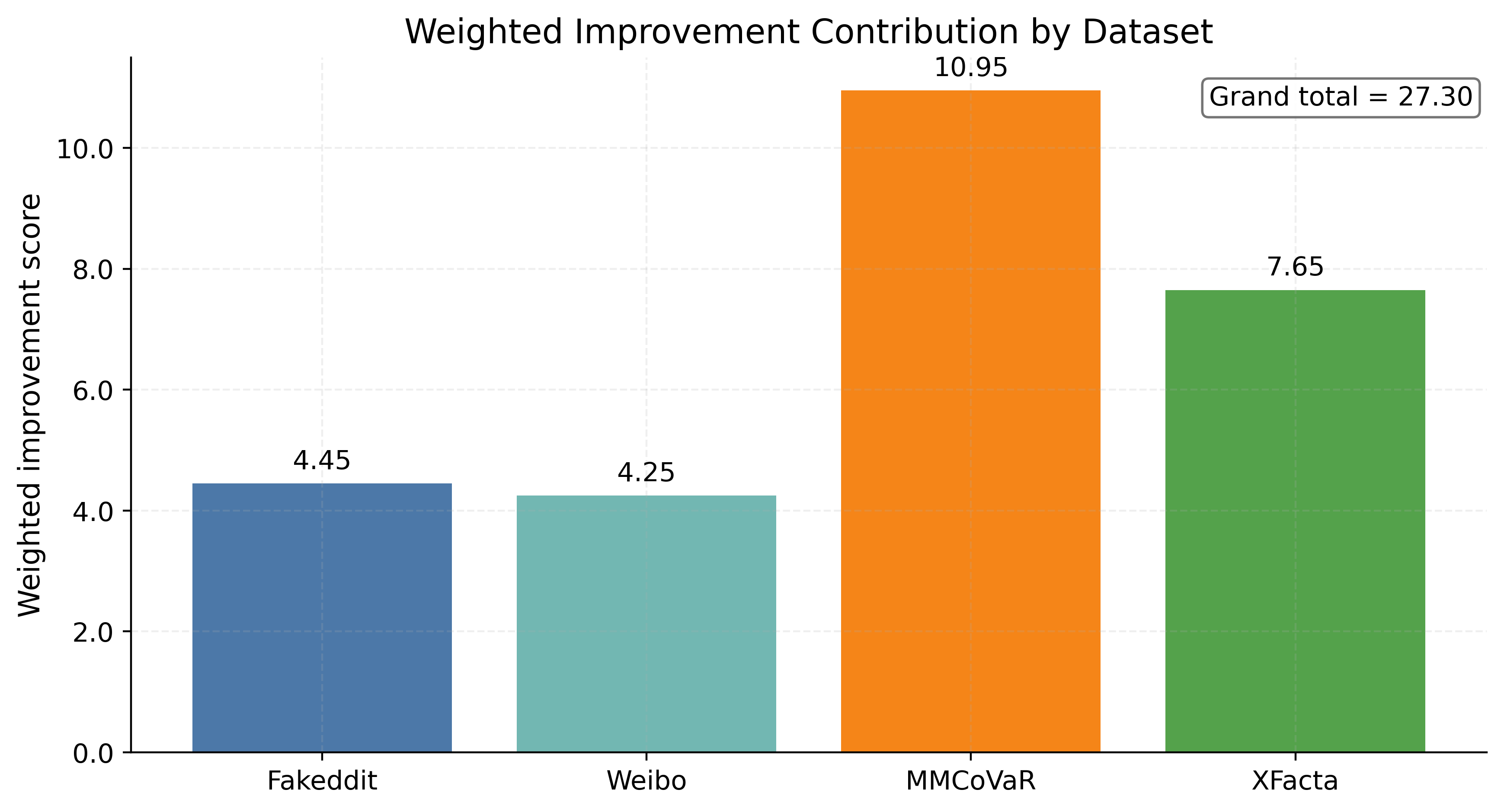}}
\caption{Weighted improvement contribution by dataset (analysis index).}
\label{fig:results_weighted_gain}
\end{figure}

Figure~\ref{fig:results_weighted_gain} shows that MOMENTA's relative gains are distributed across all four datasets rather than being concentrated in a single benchmark. The largest contributions come from the more challenging settings, especially MMCoVaR and XFacta, which aligns with the benchmark comparisons showing larger absolute improvements in these harder cross-domain scenarios.

\subsection{Discussion}
\label{sec:results_discussion}
Overall, the empirical evidence supports four conclusions. First, MOMENTA delivers consistently strong performance across heterogeneous datasets rather than overfitting to a single benchmark profile. Second, the model surpasses the strongest prior baseline in all four head-to-head comparisons, with the largest absolute gains on MMCoVaR and XFacta, where cross-domain variation is typically more pronounced. Third, diagnostic analyses show that improvements are not confined to one metric family: discrimination (AUC), class-balanced effectiveness (Macro-F1/MCC), and operating-point behavior (precision/recall/specificity under calibrated thresholds) remain coherent. Fourth, the residual-error and weighted-gain views indicate that the largest remaining headroom is concentrated in the hardest domains, which is consistent with practical expectations in real-world misinformation settings.

From a practical perspective, these results suggest that MOMENTA is suitable for deployment scenarios requiring robust behavior across multiple data sources and content regimes ~\cite{Guo2017Calibration}. The combination of expert specialization, discrepancy-aware fusion, and temporal-aware learning appears to provide both performance and stability benefits when evaluated under complementary views.

At the same time, two limitations should be acknowledged. The weighted-gain plot is a secondary descriptive index and should not be interpreted as a substitute for primary benchmark metrics. In addition, while current results demonstrate strong cross-dataset robustness, further evaluation under explicit temporal drift and out-of-distribution stress tests would strengthen external-validity claims ~\cite{XFacta,MMfakeBench}. These directions provide a clear path for future extensions of the framework.

\subsection{Code Availability}
To support transparency and reproducibility, the official implementation of \textit{MOMENTA} is available at \url{https://github.com/Yegi03/momenta}, including training and evaluation scripts.

\section{Conclusion}

In this work, we presented MOMENTA, a unified framework for multimodal misinformation detection that combines modality-specific expert specialization, cross-modal interaction, and temporal reasoning within a single architecture. By integrating Mixture-of-Experts (MoE) layers with bidirectional co-attention and an explicit discrepancy modeling branch, the framework captures both semantic alignment and inconsistency between textual and visual modalities, which are key signals in multimodal misinformation detection~\cite{ECaTCH,ooc_multimodal_misinformation,liu2025mimoe,liu2024misdmoe,CMAF}.

To address the evolving nature of misinformation, we introduced a temporal aggregation mechanism that models narrative dynamics through drift and momentum encoding over overlapping time windows. This design enables the model to capture both short-term variations and longer-term trends in content evolution, which have been shown to be important for detecting coordinated or evolving misinformation campaigns~\cite{ECaTCH,jin2025infodemics,misinformation_spread_dynamics_2025}. In addition, domain-adversarial training and prototype-based alignment were incorporated to improve cross-domain generalization, allowing the model to maintain stable performance across heterogeneous datasets~\cite{ganin2016domain,Wang2022DGSurvey}.

Experimental results on Fakeddit, MMCoVaR, Weibo, and XFacta demonstrate that MOMENTA consistently outperforms existing methods across multiple evaluation metrics. The improvements are not limited to a single metric, but extend across AUC, F1, and MCC, indicating gains in both ranking quality and class-balanced performance~\cite{Sokolova2009PerfMeasures,Chicco2020MCC}. Diagnostic analyses further show that these improvements are stable across datasets and remain robust after dataset-specific validation-based threshold calibration, suggesting that the model generalizes beyond individual benchmark settings~\cite{Guo2017Calibration,Zou2016BestThreshold}.

Overall, the results indicate that combining heterogeneous modeling, temporal reasoning, and domain-invariant representation learning provides a practical and effective direction for multimodal misinformation detection in real-world settings~\cite{XFacta,MMfakeBench}.

\section{Limitations and Future Directions}

Despite its strong performance, the proposed framework has several limitations that point to meaningful directions for future work.

First, the integration of multiple components, including Mixture-of-Experts layers, temporal aggregation, and domain-adversarial training, increases architectural complexity. While the integrated design is associated with improved overall performance, it also introduces additional computational overhead, which may limit scalability in large-scale or real-time deployment scenarios~\cite{Shazeer2017MoE,Fedus2022SwitchTransformer}. Future work should explore more efficient model designs, such as sparsely activated expert layers, parameter sharing strategies, or model compression techniques, to reduce resource requirements without sacrificing performance~\cite{Shazeer2017MoE,Fedus2022SwitchTransformer}.

Second, the temporal modeling strategy relies on predefined window sizes and fixed hyperparameters such as decay rates and momentum coefficients, and the current study does not include a dedicated sensitivity analysis for these settings. Although this formulation captures general patterns of narrative evolution, it may not fully adapt to irregular posting behavior or abrupt shifts in misinformation dynamics. As a result, long-range dependencies and non-uniform temporal structures may not be fully represented~\cite{ECaTCH,TGAT2020}. Future work may address this limitation by incorporating adaptive or continuous-time representations, as well as graph-based or sequence-level temporal models that can better capture complex evolution patterns~\cite{TGAT2020,Monti2019GeometricFakeNews}.

Third, although domain-adversarial training and prototype alignment improve cross-domain generalization, the evaluation remains limited to a finite set of benchmark datasets. Moreover, not all multimodal misinformation is characterized by overt text-image inconsistency; some cases involve mutually consistent modalities that nevertheless misrepresent external context, which may require stronger retrieval- or evidence-grounded reasoning. The model’s robustness under more extreme domain shifts, including unseen platforms, languages, or emerging misinformation formats, remains an open question~\cite{MMfakeBench,XFacta,Wang2022DGSurvey}. Future research could investigate test-time adaptation, continual learning, or meta-learning approaches to enable the model to adapt to evolving distributions without requiring full retraining~\cite{Tent2021}.

Finally, the current framework focuses on text and image modalities and does not incorporate additional signals such as user interactions, network structure, or propagation dynamics. These factors have been shown to provide complementary information for misinformation detection, particularly in social media environments~\cite{Monti2019GeometricFakeNews,Shu2017DataMiningPerspective}. Extending the framework to include additional modalities, such as video, audio, and user-level metadata, as well as integrating propagation-based features, represents a promising direction for improving performance in real-world settings~\cite{LEMMA,MMfakeBench,XFacta}.

In addition to these directions, improving interpretability remains an important challenge. While the discrepancy modeling component provides some insight into cross-modal inconsistencies, further work is needed to develop more transparent and human-interpretable explanations that can support decision-making in high-stakes applications~\cite{Shang2022DGExplain,dEFEND,CMAF}.

\bibliography{ref}

@article{Kathigi2025FakeNews,
  author  = {Kathigi, A. and Pujar, M. and Akshatha, A. M. S. and Shilpa, R. and Shirabadagi, S. S.},
  title   = {Fake News Detection Using Weighted Fine-Tuned BERT and Sparse Recurrent Neural Network},
  journal = {Journal of Computer Science},
  year    = {2025},
  volume  = {21},
  number  = {12},
  pages   = {2951--2964},
  doi     = {10.3844/jcssp.2025.2951.2964}
}

@inproceedings{rajabi2024event,
  title={Event-based multi-modal fusion for online misinformation detection in high-impact events},
  author={Rajabi, Javad and Okechukwu, Sunday and Mousavi, Ahmad and Corizzo, Roberto and Cavalcante, Charles C and Boukouvalas, Zois},
  booktitle={2024 IEEE International Conference on Big Data (BigData)},
  pages={3301--3308},
  year={2024},
  organization={IEEE}
}

@misc{DCCF,
      title={Disentangling Fact from Sentiment: A Dynamic Conflict-Consensus Framework for Multimodal Fake News Detection}, 
      author={Weilin Zhou and Zonghao Ying and Junjie Mu and Shengwei Tian and Quanchen Zou and Deyue Zhang and Dongdong Yang and Xiangzheng Zhang},
      year={2025},
      eprint={2512.20670},
      archivePrefix={arXiv},
      primaryClass={cs.LG},
      url={https://arxiv.org/abs/2512.20670}, 
}

@misc{KEN,
      title={KEN: Knowledge Augmentation and Emotion Guidance Network for Multimodal Fake News Detection}, 
      author={Peican Zhu and Yubo Jing and Le Cheng and Keke Tang and Yangming Guo},
      year={2025},
      eprint={2507.09647},
      archivePrefix={arXiv},
      primaryClass={cs.MM},
      url={https://arxiv.org/abs/2507.09647}, 
}

@misc{MIMoE,
      title={Modality Interactive Mixture-of-Experts for Fake News Detection}, 
      author={Yifan Liu and Yaokun Liu and Zelin Li and Ruichen Yao and Yang Zhang and Dong Wang},
      year={2025},
      eprint={2501.12431},
      archivePrefix={arXiv},
      primaryClass={cs.LG},
      url={https://arxiv.org/abs/2501.12431}, 
}

@article{CMAF,
  author  = {Yi, Zepu and Tang, Chenxu and Lu, Songfeng},
  title   = {User Comment-Guided Cross-Modal Attention for Interpretable Multimodal Fake News Detection},
  journal = {Applied Sciences},
  year    = {2025},
  volume  = {15},
  number  = {14},
  pages   = {7904},
  doi     = {10.3390/app15147904}
}

@article{MMfakeBench,
  title={MMFakeBench: A Mixed-Source Multimodal Misinformation Detection Benchmark for LVLMs},
  author={Liu, Xuannan and Li, Zekun and Li, Peipei and Xia, Shuhan and Cui, Xing and Huang, Linzhi and Huang, Huaibo and Deng, Weihong and He, Zhaofeng},
  journal={arXiv preprint arXiv:2406.08772},
  year={2024}
}

@inproceedings{Shang2022DGExplain,
  author    = {Shang, Lanyu and Kou, Ziyi and Zhang, Yang and Wang, Dong},
  title     = {A Duo-Generative Approach to Explainable Multimodal COVID-19 Misinformation Detection},
  booktitle = {Proceedings of the ACM Web Conference 2022 (WWW '22)},
  pages     = {3623--3631},
  year      = {2022},
  doi       = {10.1145/3485447.3512257},
  url       = {https://doi.org/10.1145/3485447.3512257},
  publisher = {ACM}
}

@inproceedings{dEFEND,
  author    = {Shu, Kai and Cui, Limeng and Wang, Suhang and Lee, Dongwon and Liu, Huan},
  title     = {dEFEND: Explainable Fake News Detection},
  booktitle = {Proceedings of the 25th ACM SIGKDD International Conference on Knowledge Discovery and Data Mining (KDD '19)},
  pages     = {395--405},
  year      = {2019},
  doi       = {10.1145/3292500.3330935},
  url       = {https://doi.org/10.1145/3292500.3330935},
  publisher = {ACM}
}

@misc{XFacta,
      title={XFacta: Contemporary, Real-World Dataset and Evaluation for Multimodal Misinformation Detection with Multimodal LLMs}, 
      author={Yuzhuo Xiao and Zeyu Han and Yuhan Wang and Huaizu Jiang},
      year={2025},
      eprint={2508.09999},
      archivePrefix={arXiv},
      primaryClass={cs.CL},
      url={https://arxiv.org/abs/2508.09999}, 
}

@misc{LEMMA,
      title={LEMMA: Towards LVLM-Enhanced Multimodal Misinformation Detection with External Knowledge Augmentation}, 
      author={Keyang Xuan and Li Yi and Fan Yang and Ruochen Wu and Yi R. Fung and Heng Ji},
      year={2024},
      eprint={2402.11943},
      archivePrefix={arXiv},
      primaryClass={cs.CL},
      url={https://arxiv.org/abs/2402.11943}, 
}

@article{Fakeddit,
    title={r/Fakeddit: A New Multimodal Benchmark Dataset for Fine-grained Fake News Detection},
    author={Nakamura, Kai and Levy, Sharon and Wang, William Yang},
    journal={arXiv preprint arXiv:1911.03854},
    year={2019}
}

@misc{MMCoVaR,
      title={MMCoVaR: Multimodal COVID-19 Vaccine Focused Data Repository for Fake News Detection and a Baseline Architecture for Classification}, 
      author={Mingxuan Chen and Xinqiao Chu and K. P. Subbalakshmi},
      year={2021},
      eprint={2109.06416},
      archivePrefix={arXiv},
      primaryClass={cs.IR},
      url={https://arxiv.org/abs/2109.06416}, 
}

@article{Weibo,
  author    = {Lv, J. and Gao, Y. and Li, L. and others},
  title     = {Multi-modal fake news detection: A comprehensive survey on deep learning technology, advances, and challenges},
  journal   = {Journal of King Saud University Computer and Information Sciences},
  volume    = {37},
  pages     = {306},
  year      = {2025},
  doi       = {10.1007/s44443-025-00317-7}
}

@article{stimpson2025perceived,
  author  = {Stimpson, Jim P. and Park, Sungchul and Adhikari, Emily H. and Nelson, David B. and Ortega, Alexander N.},
  title   = {Perceived Health Misinformation on Social Media and Public Trust in Health Care},
  journal = {Medical Care},
  year    = {2025},
  volume  = {63},
  number  = {9},
  pages   = {686--693},
  doi     = {10.1097/MLR.0000000000002180}
}

@article{jmir2024healthmisinformation,
  author  = {Gaysynsky, Anna and Senft Everson, Nicole and Heley, Kathryn and Chou, Wen-Ying Sylvia},
  title   = {Perceptions of Health Misinformation on Social Media: Cross-Sectional Survey Study},
  journal = {JMIR Infodemiology},
  year    = {2024},
  volume  = {4},
  pages   = {e51127},
  doi     = {10.2196/51127}
}

@article{infodemic_review_2022,
  author  = {Borges do Nascimento, Israel J{\'u}nior and Pizarro, Ana Beatriz and Almeida, Jussara M. and Azzopardi-Muscat, Natasha and Gon{\c{c}}alves, Marcos Andr{\'e} and Bj{\"o}rklund, Maria and Novillo-Ortiz, David},
  title   = {Infodemics and Health Misinformation: A Systematic Review of Reviews},
  journal = {Bulletin of the World Health Organization},
  year    = {2022},
  volume  = {100},
  number  = {9},
  pages   = {544--561},
  doi     = {10.2471/BLT.21.287654}
}

@misc{ECaTCH,
  title         = {E-CaTCH: Event-Centric Cross-Modal Attention with Temporal Consistency and Class-Imbalance Handling for Misinformation Detection},
  author        = {Mousavi, Ahmad and Abdollahinejad, Yeganeh and Corizzo, Roberto and Japkowicz, Nathalie and Boukouvalas, Zois},
  year          = {2025},
  eprint        = {2508.11197},
  archivePrefix = {arXiv},
  primaryClass  = {cs.CL},
  url           = {https://arxiv.org/abs/2508.11197}
}

@misc{ooc_multimodal_misinformation,
  title         = {Detecting Out-of-Context Multimodal Misinformation with interpretable neural-symbolic model},
  author        = {Zhang, Yizhou and Trinh, Loc and Cao, Defu and Cui, Zijun and Liu, Yan},
  year          = {2023},
  eprint        = {2304.07633},
  archivePrefix = {arXiv},
  primaryClass  = {cs.CL},
  doi           = {10.48550/arXiv.2304.07633},
  url           = {https://arxiv.org/abs/2304.07633}
}

@inproceedings{conneau2020xlm,
  author    = {Conneau, Alexis and Khandelwal, Kartikay and Goyal, Naman and Chaudhary, Vishrav and Wenzek, Guillaume and Guzm{\'a}n, Francisco and Grave, Edouard and Ott, Myle and Zettlemoyer, Luke and Stoyanov, Veselin},
  title     = {Unsupervised Cross-lingual Representation Learning at Scale},
  booktitle = {Proceedings of the 58th Annual Meeting of the Association for Computational Linguistics},
  year      = {2020},
  pages     = {8440--8451},
  publisher = {Association for Computational Linguistics},
  doi       = {10.18653/v1/2020.acl-main.747},
  url       = {https://aclanthology.org/2020.acl-main.747/}
}

@inproceedings{radford2021learning,
  author    = {Radford, Alec and Kim, Jong Wook and Hallacy, Chris and Ramesh, Aditya and Goh, Gabriel and Agarwal, Sandhini and Sastry, Girish and Askell, Amanda and Mishkin, Pamela and Clark, Jack and Krueger, Gretchen and Sutskever, Ilya},
  title     = {Learning Transferable Visual Models From Natural Language Supervision},
  booktitle = {Proceedings of the 38th International Conference on Machine Learning},
  year      = {2021},
  volume    = {139},
  series    = {Proceedings of Machine Learning Research},
  pages     = {8748--8763},
  publisher = {PMLR},
  url       = {https://proceedings.mlr.press/v139/radford21a.html}
}

@article{ganin2016domain,
  author  = {Ganin, Yaroslav and Ustinova, Evgeniya and Ajakan, Hana and Germain, Pascal and Larochelle, Hugo and Laviolette, Fran{\c{c}}ois and Marchand, Mario and Lempitsky, Victor},
  title   = {Domain-Adversarial Training of Neural Networks},
  journal = {Journal of Machine Learning Research},
  year    = {2016},
  volume  = {17},
  number  = {59},
  pages   = {1--35},
  url     = {https://jmlr.org/papers/v17/15-239.html}
}

@inproceedings{liu2025mimoe,
  author    = {Liu, Yifan and Liu, Yaokun and Li, Zelin and Yao, Ruichen and Zhang, Yang and Wang, Dong},
  title     = {Modality Interactive Mixture-of-Experts for Fake News Detection},
  booktitle = {Proceedings of the ACM Web Conference 2025},
  year      = {2025},
  publisher = {ACM},
  doi       = {10.1145/3696410.3714522},
  url       = {https://doi.org/10.1145/3696410.3714522}
}

@inproceedings{liu2024misdmoe,
  author    = {Liu, Moyang and Yan, Kaiying and Liu, Yukun and Fu, Ruibo and Wen, Zhengqi and Liu, Xuefei and Li, Chenxing},
  title     = {MisD-MoE: A Multimodal Misinformation Detection Framework with Adaptive Feature Selection},
  booktitle = {Proceedings of The 4th NeurIPS Efficient Natural Language and Speech Processing Workshop},
  year      = {2024},
  volume    = {262},
  series    = {Proceedings of Machine Learning Research},
  pages     = {114--122},
  publisher = {PMLR},
  url       = {https://proceedings.mlr.press/v262/liu24a.html}
}

@inproceedings{lin2017focal,
  author    = {Lin, Tsung-Yi and Goyal, Priya and Girshick, Ross and He, Kaiming and Doll{\'a}r, Piotr},
  title     = {Focal Loss for Dense Object Detection},
  booktitle = {Proceedings of the IEEE International Conference on Computer Vision (ICCV)},
  year      = {2017},
  pages     = {2980--2988},
  doi       = {10.1109/ICCV.2017.324},
  url       = {https://openaccess.thecvf.com/content_ICCV_2017/html/Lin_Focal_Loss_for_ICCV_2017_paper.html}
}

@inproceedings{liang2021rdrop,
  author    = {Liang, Xiaobo and Wu, Lijun and Li, Juntao and Wang, Yue and Meng, Qi and Qin, Tao and Chen, Wei and Zhang, Min and Liu, Tie-Yan},
  title     = {R-Drop: Regularized Dropout for Neural Networks},
  booktitle = {Advances in Neural Information Processing Systems},
  year      = {2021},
  volume    = {34},
  pages     = {10890--10905},
  url       = {https://openreview.net/forum?id=bw5Arp3O3eY}
}

@article{jin2025infodemics,
  author  = {John, J. N. and Gorman, Sara E. and Larson, Heidi J. and Jamieson, Kathleen Hall},
  title   = {Understanding Interventions to Address Infodemics Through Epidemiological, Socioecological, and Environmental Health Models: Framework Analysis},
  journal = {JMIR Infodemiology},
  year    = {2025},
  volume  = {5},
  pages   = {e67119},
  doi     = {10.2196/67119},
  url     = {https://infodemiology.jmir.org/2025/1/e67119}
}

@article{misinformation_spread_dynamics_2025,
  author  = {Jain, K. and Achuthan, K.},
  title   = {Modeling the Dynamics of Misinformation Spread: A Multi-Scenario Analysis Incorporating User Awareness and Generative AI Impact},
  journal = {Frontiers in Computer Science},
  year    = {2025},
  volume  = {7},
  pages   = {1570085},
  doi     = {10.3389/fcomp.2025.1570085},
  url     = {https://www.frontiersin.org/journals/computer-science/articles/10.3389/fcomp.2025.1570085/full}
}

@article{assessing_misinformation_infectious_diseases,
  author  = {Bernardin, Alejandro and Perez-Acle, Tomas},
  title   = {Assessing the Impact of Misinformation During the Spread of Infectious Diseases},
  journal = {Scientific Reports},
  year    = {2025},
  volume  = {15},
  pages   = {34740},
  doi     = {10.1038/s41598-025-18457-1},
  url     = {https://www.nature.com/articles/s41598-025-18457-1}
}

@article{Sokolova2009PerfMeasures,
  author  = {Sokolova, Marina and Lapalme, Guy},
  title   = {A Systematic Analysis of Performance Measures for Classification Tasks},
  journal = {Information Processing \& Management},
  year    = {2009},
  volume  = {45},
  number  = {4},
  pages   = {427--437},
  doi     = {10.1016/j.ipm.2009.03.002}
}

@article{Chicco2020MCC,
  author  = {Chicco, Davide and Jurman, Giuseppe},
  title   = {The Advantages of the Matthews Correlation Coefficient (MCC) over F1 Score and Accuracy in Binary Classification Evaluation},
  journal = {BMC Genomics},
  year    = {2020},
  volume  = {21},
  number  = {1},
  pages   = {6},
  doi     = {10.1186/s12864-019-6413-7}
}

@article{Saito2015PRROC,
  author  = {Saito, Takaya and Rehmsmeier, Marc},
  title   = {The Precision-Recall Plot Is More Informative than the ROC Plot When Evaluating Binary Classifiers on Imbalanced Datasets},
  journal = {PLOS ONE},
  year    = {2015},
  volume  = {10},
  number  = {3},
  pages   = {e0118432},
  doi     = {10.1371/journal.pone.0118432}
}

@inproceedings{Guo2017Calibration,
  author    = {Guo, Chuan and Pleiss, Geoff and Sun, Yu and Weinberger, Kilian Q.},
  title     = {On Calibration of Modern Neural Networks},
  booktitle = {Proceedings of the 34th International Conference on Machine Learning},
  series    = {Proceedings of Machine Learning Research},
  volume    = {70},
  pages     = {1321--1330},
  year      = {2017},
  publisher = {PMLR},
  url       = {https://proceedings.mlr.press/v70/guo17a.html}
}

@article{Zou2016BestThreshold,
  author  = {Zou, Quan and Xie, Shihe and Lin, Zhen and Wu, Meijuan and Ju, Yizhou},
  title   = {Finding the Best Classification Threshold in Imbalanced Classification},
  journal = {Big Data Research},
  year    = {2016},
  volume  = {5},
  pages   = {2--8},
  doi     = {10.1016/j.bdr.2015.12.001}
}

@article{Wang2022DGSurvey,
  author  = {Wang, Jindong and Lan, Cuiling and Liu, Chang and Ouyang, Yidong and Qin, Tao and Lu, Wang and Chen, Yiqiang and Zeng, Wenjun and Yu, Philip S.},
  title   = {Generalizing to Unseen Domains: A Survey on Domain Generalization},
  journal = {IEEE Transactions on Knowledge and Data Engineering},
  year    = {2023},
  volume  = {35},
  number  = {8},
  pages   = {8052--8072},
  doi     = {10.1109/TKDE.2022.3178128},
  url     = {https://doi.org/10.1109/TKDE.2022.3178128}
}

@inproceedings{Tent2021,
  author    = {Wang, Dequan and Shelhamer, Evan and Liu, Shaoteng and Olshausen, Bruno and Darrell, Trevor},
  title     = {Tent: Fully Test-Time Adaptation by Entropy Minimization},
  booktitle = {International Conference on Learning Representations},
  year      = {2021},
  url       = {https://openreview.net/forum?id=uXl3bZLkr3c}
}

@article{Shazeer2017MoE,
  author  = {Shazeer, Noam and Mirhoseini, Azalia and Maziarz, Krzysztof and Davis, Andy and Le, Quoc and Hinton, Geoffrey and Dean, Jeff},
  title   = {Outrageously Large Neural Networks: The Sparsely-Gated Mixture-of-Experts Layer},
  journal = {arXiv preprint arXiv:1701.06538},
  year    = {2017},
  url     = {https://arxiv.org/abs/1701.06538}
}

@article{Fedus2022SwitchTransformer,
  author  = {Fedus, William and Zoph, Barret and Shazeer, Noam},
  title   = {Switch Transformers: Scaling to Trillion Parameter Models with Simple and Efficient Sparsity},
  journal = {Journal of Machine Learning Research},
  year    = {2022},
  volume  = {23},
  number  = {120},
  pages   = {1--39},
  url     = {https://jmlr.org/papers/v23/21-0998.html}
}

@inproceedings{TGAT2020,
  author    = {Xu, Da and Ruan, Chuanwei and Korpeoglu, Evren and Kumar, Sushant and Achan, Kannan},
  title     = {Inductive Representation Learning on Temporal Graphs},
  booktitle = {International Conference on Learning Representations},
  year      = {2020},
  url       = {https://openreview.net/forum?id=rJeW1yHYwH}
}

@inproceedings{Monti2019GeometricFakeNews,
  author    = {Monti, Federico and Frasca, Fabrizio and Eynard, Davide and Mannion, Damon and Bronstein, Michael M.},
  title     = {Fake News Detection on Social Media using Geometric Deep Learning},
  booktitle = {International Conference on Learning Representations},
  year      = {2019},
  url       = {https://openreview.net/forum?id=HJgcvJQK8m}
}

@article{Shu2017DataMiningPerspective,
  author  = {Shu, Kai and Sliva, Amy and Wang, Suhang and Tang, Jiliang and Liu, Huan},
  title   = {Fake News Detection on Social Media: A Data Mining Perspective},
  journal = {ACM SIGKDD Explorations Newsletter},
  year    = {2017},
  volume  = {19},
  number  = {1},
  pages   = {22--36},
  doi     = {10.1145/3137597.3137600},
  url     = {https://doi.org/10.1145/3137597.3137600}
}

@article{acemoglu2010misinformation,
  author  = {Acemoglu, Daron and Ozdaglar, Asuman and ParandehGheibi, Ali},
  title   = {Spread of (mis)information in social networks},
  journal = {Games and Economic Behavior},
  year    = {2010},
  volume  = {70},
  number  = {2},
  pages   = {194--227}
}

@book{floridi2011information,
  author    = {Floridi, Luciano},
  title     = {The Philosophy of Information},
  year      = {2011},
  publisher = {Oxford University Press},
  address   = {Oxford}
}

@incollection{lackey2021echos,
  author    = {Lackey, Jennifer},
  title     = {Echo chambers, fake news, and social epistemology},
  booktitle = {The Epistemology of Fake News},
  editor    = {Bernecker, S. and Flowerree, A. K. and Grundmann, T.},
  publisher = {Oxford University Press},
  address   = {New York},
  year      = {2021}
}

@article{lewandowsky2023misinformation,
  author  = {Lewandowsky, Stephan and Ecker, Ullrich K. H. and Cook, John and Van Der Linden, Stephan and Roozenbeek, Jon and Oreskes, Naomi},
  title   = {Misinformation and the epistemic integrity of democracy},
  journal = {Current Opinion in Psychology},
  year    = {2023},
  volume  = {54},
  pages   = {101711}
}

@inproceedings{Naeemul,
    title = "Not all Fake News is Written: A Dataset and Analysis of Misleading Video Headlines",
    author = "Sung, Yoo Yeon  and
      Boyd-Graber, Jordan  and
      Hassan, Naeemul",
    editor = "Bouamor, Houda  and
      Pino, Juan  and
      Bali, Kalika",
    booktitle = "Proceedings of the 2023 Conference on Empirical Methods in Natural Language Processing",
    month = dec,
    year = "2023",
    address = "Singapore",
    publisher = "Association for Computational Linguistics",
    url = "https://aclanthology.org/2023.emnlp-main.1010/",
    doi = "10.18653/v1/2023.emnlp-main.1010",
    pages = "16241--16258",
}

@article{zhang2025kamp,
  title   = {Knowledge-aware multimodal pre-training for fake news detection},
  author  = {Zhang, Litian and Zhang, Xiaoming and Zhou, Ziyi and Zhang, Xi and Yu, Philip S. and Li, Chaozhuo},
  journal = {Information Fusion},
  volume  = {114},
  pages   = {102715},
  year    = {2025},
  issn    = {1566-2535},
  doi     = {10.1016/j.inffus.2024.102715}
}

\end{document}